\documentclass{aa}  

\usepackage{graphicx}
\usepackage{txfonts}
\usepackage{booktabs}
\usepackage[dvipsnames]{xcolor}
\usepackage{colortbl}
\usepackage{threeparttable}
\usepackage{soul}
\usepackage{mathtools}
\usepackage{hyperref}
\hypersetup{
    colorlinks=true,
    linkcolor=blue,
    citecolor=blue,
    filecolor=magenta,      
    urlcolor=cyan,
    }

\newcolumntype{a}{>{\columncolor{OrangeRed}}c}
\setstcolor{OrangeRed}

\begin{document}
\newcommand{\hi}{\textsc{H$\,$i}}
\newcommand{\htwo}{\ensuremath{\mathrm{H_2}}}

   \title{VERTICO VI: Cold-gas asymmetries in Virgo cluster galaxies}


\author{Ian D. Roberts\inst{\ref{leiden}}
        \and
        Toby Brown\inst{\ref{herzberg}}
        \and
        Nikki Zabel\inst{\ref{capetown}}
        \and
        Christine D. Wilson\inst{\ref{mcmaster}}
        \and
        Aeree Chung\inst{\ref{yonsei}}
        \and
        Laura C. Parker\inst{\ref{mcmaster}}        
        \and
        Dhruv Bisaria\inst{\ref{queens}}
        \and
        Alessandro Boselli\inst{\ref{marseille}}
        \and
        Barbara Catinella\inst{\ref{icrar},\ref{astro3d}}
        \and
        Ryan Chown\inst{\ref{western},\ref{western_earth_space}}
        \and
        Luca Cortese\inst{\ref{icrar},\ref{astro3d}}
        \and
        Timothy A. Davis\inst{\ref{cardiff}}
        \and
        Sara Ellison\inst{\ref{victoria}}
        \and
        Mar\'{i}a Jes\'{u}s Jim\'{e}nez-Donaire\inst{\ref{observatorio_nacional},\ref{observatorio_yebes}}
        \and
        Bumhyun Lee\inst{\ref{kassi}}
        \and
        Rory Smith\inst{\ref{chile}}
        \and
        Kristine Spekkens\inst{\ref{rmc}}
        \and
        Adam R. H. Stevens\inst{\ref{icrar}}
        \and
        Mallory Thorp\inst{\ref{victoria}}
        \and
        Vicente Villanueva\inst{\ref{maryland}}
        \and
        Adam B. Watts\inst{\ref{icrar},\ref{astro3d}}
        \and
        Charlotte Welker\inst{\ref{jhu}}
        \and
        Hyein Yoon\inst{\ref{astro3d},\ref{sydney}}
       }

\institute{Leiden Observatory, Leiden University, PO Box 9513, 2300 RA
Leiden, The Netherlands \label{leiden}
\\\email{iroberts@strw.leidenuniv.nl}
\and
Herzberg Astronomy and Astrophysics Research Centre, National Research Council of Canada, 5071 West Saanich Road, Victoria, BC 8 V9E 2E7, Canada \label{herzberg}
\and
Department of Astronomy, University of Cape Town, Private Bag X3, Rondebosch 7701, South Africa \label{capetown}
\and
Department of Physics \& Astronomy, McMaster University, 1280 Main Street W, Hamilton, ON L8S 4M1, Canada \label{mcmaster}
\and
Department of Astronomy, Yonsei University, 50 Yonsei-ro, Seodaemun-gu, Seoul, 03722, Republic of Korea \label{yonsei}
\and
Department of Physics, Engineering Physics and Astronomy, Queen’s University, Kingston, ON K7L 3N6, Canada \label{queens}
\and
Aix Marseille Univ, CNRS, CNES, LAM, Marseille, F-13013 France \label{marseille}
\and
International Centre for Radio Astronomy Research, The University of Western Australia, 35 Stirling Highway, Crawley, WA 6009, Australia \label{icrar}
\and
ARC Centre of Excellence for All Sky Astrophysics in 3 Dimensions (ASTRO 3D), Australia \label{astro3d}
\and
Department of Physics \& Astronomy, The University of Western Ontario, London, ON N6A 3K7, Canada \label{western}
\and
Institute for Earth and Space Exploration, The University of Western Ontario, London, ON N6A 3K7, Canada \label{western_earth_space}
\and
Cardiff Hub for Astrophysics Research \&\ Technology, School of Physics \&\ Astronomy, Cardiff University, Queens Buildings, Cardiff, CF24 3AA, UK \label{cardiff}
\and
Department of Physics \& Astronomy, University of Victoria, PO Box 1700 STN CSC, Victoria, BC V8W 2Y2, Canada \label{victoria}
\and
Observatorio Astron\'{o}mico Nacional (IGN), C/Alfonso XII, 3, E-28014 Madrid, Spain \label{observatorio_nacional}
\and
Centro de Desarrollos Tecnol\'{o}gicos, Observatorio de Yebes (IGN), E-19141 Yebes, Guadalajara, Spain \label{observatorio_yebes}
\and
Korea Astronomy and Space Science Institute, 776 Daedeokdae-ro, Daejeon 34055, Republic of Korea \label{kassi}
\and
Departamento de F\'{i}sica, Universidad T\'{e}cnica Federico Santa Mar\'{i}a, Vicu\~{n}a Mackenna 3939, San Joaqu\'{i}n, Santiago de Chile, Chile \label{chile}
\and
Royal Military College of Canada, P.O. Box 17000, Station Forces, Kingston, ON, K7K 7B4, Canada \label{rmc}
\and
Department of Astronomy, University of Maryland, College Park, MD 20742, USA \label{maryland}
\and
Department of Physics \& Astronomy, Johns Hopkins University, Baltimore, MD 21218, USA \label{jhu}
\and
Sydney Institute for Astronomy, School of Physics, University of Sydney, NSW 2006, Australia \label{sydney}
}


 
\abstract{We analyze cold-gas distributions in Virgo cluster galaxies using resolved CO(2-1) (tracing molecular hydrogen, \htwo{}) and \hi{} observations from the Virgo Environment Traced In CO (VERTICO) and VLA Imaging of Virgo in Atomic Gas (VIVA) surveys. From a theoretical perspective, it is expected that environmental processes in clusters will have a stronger influence on diffuse atomic gas compared to the relatively dense molecular gas component, and that these environmental perturbations can compress the cold interstellar medium in cluster galaxies leading to elevated star formation. In this work we observationally test these predictions for star-forming satellite galaxies within the Virgo cluster. We divide our Virgo galaxy sample into \hi{}-normal, \hi{}-tailed, and \hi{}-truncated classes and show, unsurprisingly, that the \hi{}-tailed galaxies have the largest quantitative \hi{} asymmetries. We also compare to a control sample of non-cluster galaxies and find that Virgo galaxies, on average, have \hi{} asymmetries that are $40 \pm 10$ per cent larger than the control. There is less separation between control, \hi{}-normal, \hi{}-tailed, and \hi{}-truncated galaxies in terms of \htwo{} asymmetries, and on average, Virgo galaxies have \htwo{} asymmetries that are only marginally ($20 \pm 10$ per cent) larger than the control sample. We find a weak correlation between \hi{} and \htwo{} asymmetries over our entire sample, but a stronger correlation for those specific galaxies being strongly impacted by environmental perturbations. Finally, we divide the discs of the \hi{}-tailed Virgo galaxies into a leading half and trailing half according to the observed tail direction. We find evidence for excess molecular gas mass on the leading halves of the disc. This excess molecular gas on the leading half is accompanied by an excess in star formation rate such that the depletion time is, on average, unchanged.}


\maketitle
%

\section{Introduction} \label{sec:intro}

Galaxy clusters are the most massive gravitationally bound objects in the Universe and they have a strong impact on the properties of galaxies that they host. Most convincing is the increased likelihood of satellite galaxies\footnote{Here, and throughout the paper, we use the terminology `satellite galaxy' to refer to member galaxies in a cluster that are not the central brightest cluster galaxy.} in clusters to be red and passive (in terms of ongoing star formation) relative to galaxies of a similar mass but not part of an over-dense environment \citep[e.g.][]{dressler1980,postman1984,croton2005,peng2010,wetzel2012,lin2014,brown2017,jian2018,davies2019,roberts2019}. This environmental quenching is likely a product of two aspects of galaxy clusters: First, the crowded cluster field (particularly near the cluster centre) increases the number of galaxy-galaxy interactions, including both tidal effects \citep[e.g.][]{mayer2006,chung2007} and impulsive galaxy fly-by encounters (sometimes referred to as `galaxy harassment', e.g. \citealt{moore1996}). Second, the deep potential well of galaxy clusters leads to a distribution of hot, ionized gas permeating the space between galaxies, known as the intra-cluster medium (ICM). Galaxies moving through the ICM at high speed experience a ram pressure \citep[e.g.][]{gunn1972,quilis2000} that is capable of removing gas. Ram-pressure stripping (RPS) is likely an important driver of star formation quenching in dense environments (see \citealt{cortese2021,boselli2022_review} for recent reviews). This is particularly true for massive galaxy clusters, whereas the efficacy of RPS in lower-mass galaxy groups is more of an open question \citep[e.g.][]{rasmussen2006,vulcani2018_grp,kleiner2021,roberts2021_LOFARgrp,lee2022,roberts2022_UNIONS}. Quenching may manifest from a modest ram pressure that is able to remove \hi{} and the circumgalactic medium (CGM) but leaves the galaxy molecular gas component in place \citep[e.g.][]{stevens2017}. In this case (often called `starvation'), the galaxy is not able to replenish its gas reservoir and will quench on a timescale set by the depletion time of the remaining gas ($\sim$Gyr timescales, e.g.\ \citealt{bigiel2011,saintonge2017}). A more rapid quenching will occur if the ram pressure is strong enough to directly remove molecular gas from the disc, leading to a reduction in star formation on shorter timescales ($<\mathrm{Gyr}$, e.g.\ \citealt{quilis2000,pappalardo2010,boselli2016_galex,fossati2018}).
\par
A complete understanding of the impact of environment on star formation requires observational constraints on all components of the interstellar medium (ISM) in satellite galaxies, including ionized, atomic, and molecular gas. Ionized and atomic gas have been fairly widely surveyed and resolved in nearby clusters over the past decade or two \citep[e.g.][]{bravo-alfaro2000,gavazzi2001,kenney2004,chung2009,kenney2014,yagi2010,poggianti2017,boselli2018,molnar2022,hess2022}. The results of these surveys have shown that cluster galaxies tend to be deficient in \hi{}, with star-forming discs that have been truncated from the outside in \citep[e.g.][]{koopmann2004,chung2009,cortese2012,boselli2015,schaefer2017,yoon2017,schaefer2019}. Some cluster galaxies also show long, one-sided gas tails extending beyond their discs (sometimes referred to as `jellyfish galaxies', e.g. \citealt{gavazzi1987,yoshida2002,cortese2007,chung2007,chung2009,poggianti2017,boselli2016_ngc4569,boselli2018,roberts2021_LOFARclust}). These tailed galaxies are most often associated with ongoing RPS.
\par
There has been comparatively less work focused on the molecular hydrogen (\htwo{}) component of cluster galaxies, though this is rapidly changing, in large part due to sensitive molecular gas observations from the Atacama Large Millimeter/submillimeter Array (ALMA). This includes both case-study observations focused on individual, or a small number of, cluster galaxies \citep[e.g.][]{vollmer2008,vollmer2009,vollmer2012,jachym2014,verdugo2015,jachym2017,lee2017,lee2018,moretti2018,jachym2019,cramer2020,moretti2020b,moretti2020,cramer2021,roberts2022_lofar_manga} as well as surveys covering tens of galaxies within individual clusters \citep[e.g.][]{kenney1986,kenney1988,boselli1997,boselli2014,mok2016,mok2017,zabel2019,brown2021,morokuma-matsui2022,villanueva2022,zabel2022,watts2023}. Generally speaking, \htwo{} appears less perturbed by environment than \hi{} gas. Many cluster galaxies have similar \htwo{} masses relative to comparable non-cluster galaxies, but in turn are substantially more \hi{} deficient, leading to elevated molecular-to-atomic gas ratios \citep[e.g.][]{boselli2014,mok2016,loni2021,zabel2022}. Though we note that some studies have found \htwo{}-deficient cluster galaxies as well as cluster galaxies with significant \hi{} reservoirs but no detectable CO emission \citep{zabel2019,loni2021}. Elevated molecular-to-atomic gas ratios in cluster galaxies are likely a product of outside-in gas stripping where the more extended, diffuse atomic gas is removed more easily than the dense, more centrally concentrated molecular gas (see e.g.\ \citealt{stevens2021}). There is some evidence that molecular gas is also directly stripped from cluster galaxies, just not to the extent of \hi{} \citep[e.g.][]{fumagalli2009,boselli2014,zabel2019,cramer2020,watts2023}. Previous studies have detected CO emission extending beyond the optical disc in some cluster galaxies \citep{jachym2014,jachym2017,jachym2019,zabel2019,moretti2020b,moretti2020}, though there is still debate around whether this molecular gas is directly removed from the galaxy or if it is formed in situ within the tail from stripped atomic gas.
\par
This work is the latest in a series of early science papers from the Virgo Environment Traced In CO (VERTICO, \citealt{brown2021}) survey, which probe the effects of the cluster environment on molecular gas in star-forming galaxies. Previous VERTICO science papers have found: a systematic decrease in molecular gas content for \hi{}-deficient galaxies \citep{zabel2022,watts2023}, that VERTICO galaxies have more centrally-concentrated molecular gas discs than galaxies in the field \citep{zabel2022}, and that galaxies with perturbed and deficient \hi{} distributions tend to have long molecular gas depletion times \citep{villanueva2022,jimenez-donaire2022}. On the whole, these works show that there is a measurable impact from environment on molecular gas properties in Virgo star-forming galaxies.
\par
In this paper, we continue the study of molecular gas properties in Virgo satellite galaxies with VERTICO. We leverage the synergy between the VLA Imaging of Virgo in Atomic Gas (VIVA, \citealt{chung2009}) and VERTICO in order to constrain the differential effects of the cluster environment on atomic and molecular gas. We measure and contrast the quantitative asymmetries of atomic and molecular gas for Virgo star-forming galaxies as well as a control sample of non-cluster galaxies drawn from the HERA\footnote{Heterodyne Receiver Array} CO(2-1) Legacy Survey (HERACLES, \citealt{leroy2009}) and The \hi{} Nearby Galaxy Survey (THINGS, \citealt{walter2008}). For Virgo galaxies with \hi{} tails observed from VIVA, we also constrain any anisotropies between cold-gas distributions on the leading half (i.e.\ opposite to the tail direction) and the trailing half (i.e.\ in the direction of the tail). The majority of these tails are believed to be a product of RPS \citep{chung2007,yoon2017,boselli2022_review}, therefore this allows us to test whether ram pressure leads to elevated gas densities on the leading half of Virgo galaxies. This effect has been previously seen in hydrodynamical simulations \citep[e.g.][]{tonnesen2009,steinhauser2012,bekki2014,troncoso-iribarren2020} and select observational works \citep[e.g.][]{lee2017,cramer2020,boselli2021_ic3476,cramer2021,roberts2022_lofar_manga}.
\par
The structure of this paper is as follows: Section~\ref{sec:data_methods} describes the data products used in this work, and outlines the method that we use to quantify gas asymmetries. Section~\ref{sec:virgo_asym} presents a comparison between atomic and molecular gas asymmetries in Virgo galaxies and our control sample. Section~\ref{sec:mol_dense_lt} tests whether tailed galaxies show excess molecular gas emission on their leading halves. Section~\ref{sec:conclusion} provides a summary of the primary results and a brief discussion. Throughout this paper we assume a common distance of $16.5\,\mathrm{Mpc}$ to all Virgo galaxies \citep{mei2007} and a Chabrier initial mass function \citep{chabrier2003}.

\begin{figure*}
    \centering
    \includegraphics[width=\textwidth]{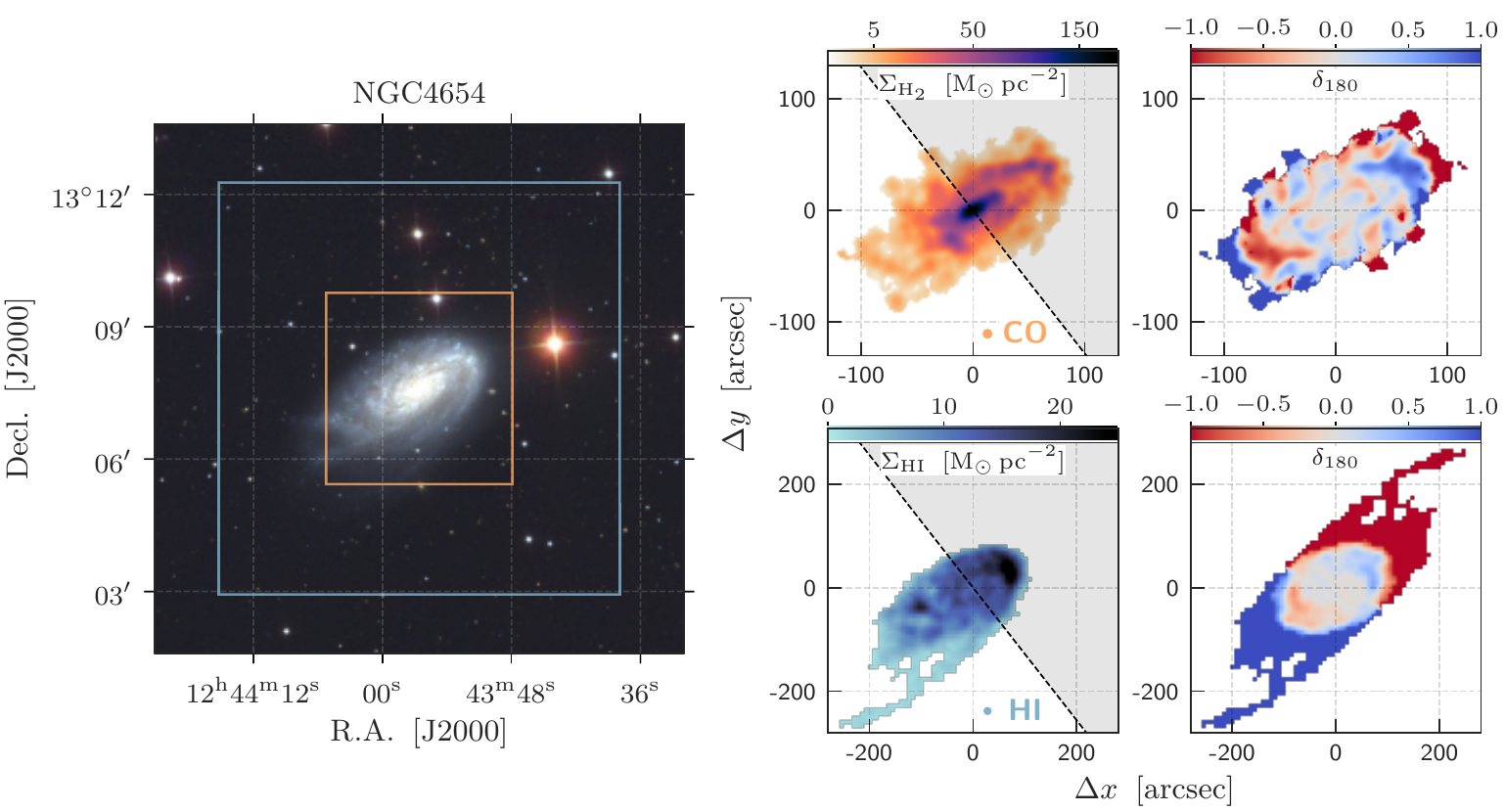}
    \caption{Example images for NGC4654. \textit{Left:} Optical $gri$ image of NGC4654 from the Sloan Digital Sky Survey \citep[e.g.][]{york2000,blanton2017}. The orange box corresponds to the CO image area and the blue box corresponds to the \hi{} image area. \textit{Right:} Gas surface density maps and asymmetry difference maps for the VERTICO CO image (top, $9\arcsec$ beam) and the VIVA \hi{} image (bottom, $16.9\arcsec \times 16.7\arcsec$ beam).  Overlaid on the gas surface density maps we show the division between the leading half (shaded) and trailing half (not shaded) for NGC4654 as used in Section~\ref{sec:mol_dense_lt} \citep{lee2022_virgo_tails}.}
    \label{fig:example_img}
\end{figure*}

\section{Data and methods} \label{sec:data_methods}

\subsection{The Virgo galaxy sample} \label{sec:data_virgo_gals}

Our parent sample of Virgo galaxies consists of 51 galaxies that are part of both VIVA \citep{chung2009} and VERTICO \citep{brown2021}. These galaxies cover a wide range in stellar mass ($M_\bigstar \sim 10^9 - 10^{11}\,\mathrm{M_\odot}$) and specific star formation rate ($\mathrm{sSFR} \equiv \mathrm{SFR} / M_\bigstar$, $10^{-11.5} - 10^{-9.5}\,\mathrm{yr^{-1}}$), and also span cluster-centric distances ranging from the centre of Virgo ($<0.5\,R_{200}$) to the outskirts ($\sim\!2.5\,R_{200}$). From these 51 galaxies we preemptively exclude five from our sample for this work: two that are $^{12}$CO (2-1) non-detections in VERTICO (IC3418 and VCC1581), the interacting NGC4567/8 system, and NGC4533 which is not included in the \citet{yoon2017} classifications that are used throughout this work. Thus our initial Virgo sample consists of 46 galaxies (see Table~\ref{tab:galaxy_table}), all of which have measured $^{12}$CO (2-1) and \hi{} emission. In Section~\ref{sec:methods_asym} we describe further cuts which we apply to this initial Virgo sample in order to ensure reliable asymmetry measurements for that part of our analysis.

\subsubsection{VERTICO molecular gas data} \label{sec:data_vertico}

VERTICO \citep{brown2021} is an ALMA large program that uses the Atacama Compact Array (ACA) to obtain kpc-scale imaging of the $^{12}$CO (2-1), hereafter CO, line for the aforementioned 51 Virgo cluster galaxies.  Of these galaxies, 15 already had archival CO observations from the ACA (2015.1.00956.S: PI A.\ Leroy, 2016.1.00912.S: PI J.\ Kenney, 2017.1.00886.L: PI E.\ Schinnerer) and the remaining 36 were observed as part of the VERTICO program in ALMA Cycle 7. Of the 36 VERTICO galaxies observed in ALMA Cycle 7, total power (TP) observations were added for 25 galaxies that were expected to have emission on scales larger than the largest angular scale of $29\arcsec$ for the ACA in ALMA Band 6. TP data was also available for 14/15 of the archival VERTICO galaxies.  VERTICO also observed the $^{13}$CO (2-1) and C$^{18}$O (2-1) lines along with the radio continuum in ALMA Band 6, but this work only uses the $^{12}$CO products. 
\par
CO was imaged using the PHANGS-ALMA imaging pipeline \citep{leroy2021_pipeline}, with three small modifications that are described in \citet{brown2021}, in order to optimize the imaging procedure for VERTICO data.  Briggs weighting \citep{briggs1995} was used with a robust parameter of 0.5 and the target velocity resolution was set to be $10\,\mathrm{km\,s^{-1}}$. For galaxies with both ACA and TP data, the two were combined via feathering using the PHANGS-ALMA pipeline \citep{leroy2021_pipeline}. The native resolution for the resulting CO cubes ranges between $7.1\arcsec$ and $10.2\arcsec$. In order to obtain uniform spatial resolution across the sample, the native resolution cubes are subsequently smoothed to both $9\arcsec$ and $15\arcsec$ resolution using \texttt{imsmooth} in CASA. At the distance of the Virgo cluster ($16.5\,\mathrm{Mpc}$), the $9\arcsec$ and $15\arcsec$ angular resolutions correspond physical scales of $720\,\mathrm{pc}$ and $1.2\,\mathrm{kpc}$, respectively. For one galaxy, NGC4321, the native resolution is greater than $9\arcsec$ (native resolution, $10.2\arcsec = 816\,\mathrm{pc}$) and thus only $10.2\arcsec$ and $15\arcsec$ products are available. For completeness we use the $816\text{-}\mathrm{pc}$ resolution products for NGC4321 alongside the $720\text{-}\mathrm{pc}$ resolution products for the rest of the Virgo sample.  Finally, detection masks and integrated intensity maps are produced for each galaxy from these reduced cubes following the procedure that is outlined in \citet{brown2021}.  Molecular gas surface density maps are made assuming $\alpha_\mathrm{CO} = 4.35\,\mathrm{M_\odot\,pc^{-2}\,(K\,km\,s^{-1})^{-1}}$ and a $\mathrm{CO\,(2\text{-}1)}$-to-$\mathrm{CO\,(1\text{-}0)}$ ratio of $R_{21} = 0.8$. This assumed value of $R_{21} = 0.8$ is consistent with the average value from the xCOLD GASS survey \citep{saintonge2017}, as well as the average value measured from 35 VERTICO galaxies with high-quality CO(1-0) data \citep{boselli2014_co,brown2021}. In Fig.~\ref{fig:example_img} we show an example of a VERTICO \htwo{} gas surface density map for NGC4654.

\subsubsection{VIVA atomic gas data} \label{sec:data_viva}

For all galaxies in our Virgo sample we use atomic gas data products from the VIVA survey. Specifically, we use reprocessed \hi{} cubes which match the resolution (as best as possible) of the $1.2\,\mathrm{kpc}$ VERTICO CO products. In order to reach this resolution, galaxies in VIVA that were observed with both the C- and D-configurations of the VLA are re-imaged only including the C-configuration data. The results are VIVA \hi{} cubes for our Virgo galaxies with synthesized beams that range between $1.2\,\mathrm{kpc}$ and $2.0\,\mathrm{kpc}$, with an average of $1.4\,\mathrm{kpc}$. From these reprocessed cubes, \hi{} detection masks and integrated intensity maps are produced with the VERTICO products pipeline, as for the CO maps. Atomic gas masses were calculated assuming optically thin emission \citep[e.g.][]{meyer2017},
\begin{equation}
    M_\mathrm{\hi{}} = 2.35 \times 10^5\,d_L^2\,(1 + z)^{-2}\,S_\mathrm{\hi{}} \quad \mathrm{[M_\odot]}
\end{equation}
\noindent
where $d_L$ is the luminosity distance to each galaxy, $z$ is the redshift, and $S_\mathrm{\hi{}}$ is the integrated flux density in $\mathrm{Jy\,km\,s^{-1}}$.
\par
An example \hi{} gas surface density map for NGC4654 is shown in Fig.~\ref{fig:example_img}. The main difference between these reprocessed VIVA products and the original VIVA images presented in \citet{chung2009} is that the increased resolution comes at a cost of the column density sensitivity and therefore diffuse flux detection for the $\sim\!10$ VIVA galaxies that were observed in D-configuration. Given the bias that can be introduced from comparing asymmetries measured at different resolutions (see Section~\ref{sec:methods_asym}), it is preferable for this analysis to have \htwo{} and \hi{} maps that are as close as possible to being resolution matched, even at the expense of some extended \hi{} emission. Furthermore, this choice not only results in roughly resolution-matched \htwo{} and \hi{} maps, but also in higher resolution \hi{} maps which also leads to more accurate asymmetry measurements (again, see Section~\ref{sec:methods_asym} for more details).  The majority of VIVA galaxies only include C-configuration data, thus by removing the D-configuration data for a minority of galaxies we ensure that the \hi{} maps for the Virgo sample all are taken with the same observing set-up.  This allows for comparisons of measured \hi{} asymmetries within the Virgo sample to be interpreted in a more straight-forward manner. Lastly, we note that even if we include the D-configuration data for these $\sim$10 VIVA galaxies in our analysis, this does not alter any of the qualitative conclusions of this work.

\subsection{The HERACLES/THINGS control sample} \label{sec:data_heracles}

\begin{figure}
    \centering
    \includegraphics{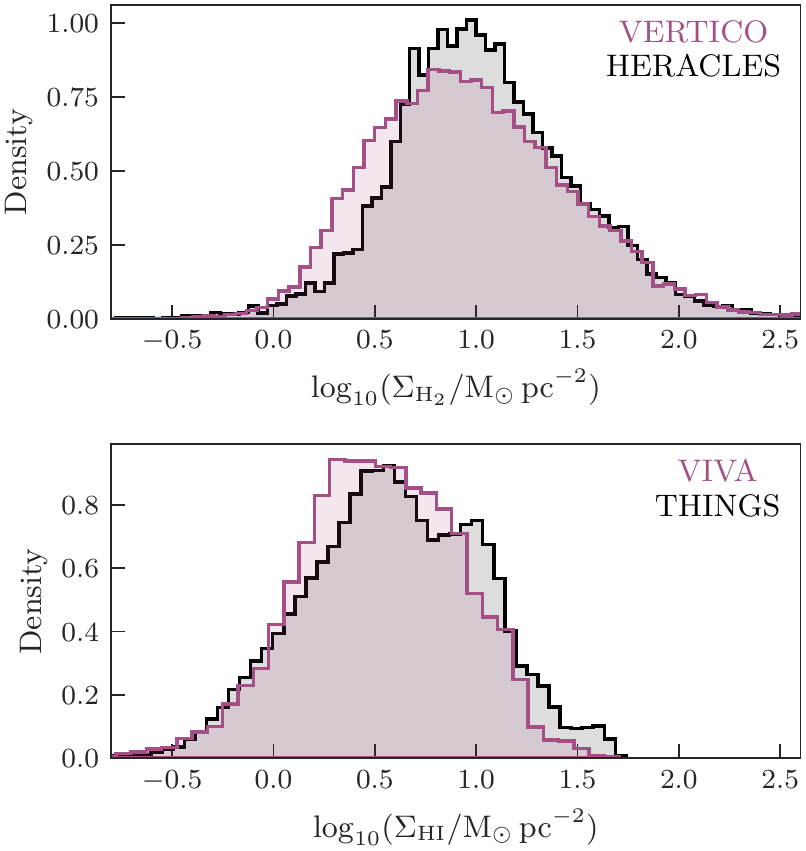}
    \caption{\htwo{} (top) and \hi{} (bottom) surface density distributions for detected pixels in the gas surface density maps of the Virgo (magenta) and control sample (black) galaxies used for the asymmetry analysis (see Table~\ref{tab:galaxy_table}). \htwo{} surface density distributions are taken from the $\mathrm{720}\text{-}\mathrm{pc}$ resolution maps and \hi{} surface density distributions are taken from the low-res maps with resolutions between $1.20$ and $2.08\,\mathrm{kpc}$ for VIVA and a common resolution of $1.4\,\mathrm{kpc}$ for THINGS. Bin widths are set according to "Knuth's rule" \citep{knuth2006} using the histogram functionality within \texttt{astropy.visualization}. The Virgo and control samples both cover similar ranges of gas surface densities, both in terms of \hi{} and \htwo{}.}
    \label{fig:gas_density_dist}
\end{figure}

For a control sample of resolved, non-cluster galaxies with atomic and molecular gas data, we use the HERACLES \citep{leroy2009} and THINGS \citep{walter2008} surveys. First we remove galaxies from HERACLES that are in the Virgo cluster (NGC4254, NGC4321, NGC4536, NGC4569, and NGC4579) as well as any interacting systems in HERACLES and THINGS (NGC2146, NGC2798, NGC3034, NGC5194, and NGC5713). Just as for VERTICO, we use two control samples at different resolutions, one matching the VERTICO products at a physical resolution of $720\,\mathrm{pc}$ and one at a physical resolution of $1.4\,\mathrm{kpc}$ in order to match the average resolution of the VIVA data. To reach these resolutions and circularize the beam, cubes were smoothed with the \texttt{imsmooth} function in CASA. Some galaxies in HERACLES/THINGS are too distant to reach $720\text{-}\mathrm{pc}$ resolution, given the native angular resolutions which are $13\arcsec$ for HERACLES and more varied for THINGS (depending on the VLA configuration). As a result, our high-resolution control sample contains fewer galaxies than our low-resolution control sample. The galaxies that make up our final resolved control sample are listed in Table~\ref{tab:galaxy_table}. Integrated intensity maps for control sample galaxies are calculated from $10\,\mathrm{km\,s^{-1}}$ cubes using the pipeline as for the VERTICO data (see \citealt{brown2021} for details).
\par
We note that the HERACLES and THINGS samples are selected from the SIRTF\footnote{The Spitzer Space Telescope, formerly the Space Infrared Telescope Facility} Nearby Galaxy Survey sample (SINGS, \citealt{kennicutt2003}), which under-represents galaxies with faint $60\text{-}\mathrm{\mu m}$ luminosities. Given the relation between infrared luminosity and gas content in star-forming galaxies, our control sample may be somewhat biased towards gas-rich galaxies, though we do not expect this to bias the measured asymmetries.

\subsection{Star formation rate and stellar mass maps} \label{sec:data_mstar}

We make use of resolved SFR and stellar mass surface density maps for our Virgo and control galaxy samples. These maps are primarily used in Section~\ref{sec:mol_dense_lt} but we also use the stellar mass maps to define the galaxy centres for the asymmetry analysis in Section~\ref{sec:virgo_asym}. Here we give a brief description of these data products, but please also see previous VERTICO publications that use these same products \citep{zabel2022,villanueva2022,jimenez-donaire2022,watts2023,brown2023}.
\par
Star formation rate maps are constructed from a combination of GALEX NUV and WISE Band-3 (WISE3, $12\,\mathrm{\mu m}$) imaging following the calibrations given by \citet{leroy2019}. Stellar mass maps are derived from \textit{WISE} Band-1 (WISE1, $3.4\,\mathrm{\mu m}$) imaging, also following \citet{leroy2019}.  For the stellar mass maps, a local mass-to-light ratio (at $3.4\,\mathrm{\mu m}$) is determined using the WISE3-to-WISE1 colour as an `sSFR-like' proxy.
\par
For all photometric products used to derive the SFR and stellar mass maps (GALEX NUV, WISE1, WISE3), we mask any \emph{Gaia} stars in the image cutouts following the procedure in \citet{leroy2019}. We then fill the mask holes in the maps by interpolating over these masked regions with the \texttt{convolve} function from \texttt{Astropy}. This gives us smooth maps and ensures that the anisotropy measurements between the leading and trailing halves in Section~\ref{sec:mol_dense_lt} are not affected by the projected distribution of foreground stars. The number of masked stars is largest for the WISE1 images, and thus the stellar mass maps. There are comparatively few UV-bright foreground stars that require masking in the SFR maps, though this is in part due to the wavelength coverage of \emph{Gaia}. Star maps are also inspected by eye, and when required, bright stars missed by this automated masking are masked by hand.
\par
We use the stellar mass map for each galaxy to define the galaxy centre about which gas asymmetries are measured (see Section~\ref{sec:methods_asym}), and for splitting \hi{}-tailed galaxies into leading and trailing halves (see Section~\ref{sec:mol_dense_lt}). We define the galaxy centre to be the centroid of the stellar mass map using the \texttt{centroids} module from the \texttt{Python} package \texttt{photutils}. Our adopted galaxy centres are determined from a `centre-of-mass' centroid estimator (\texttt{photuils.centroids} \texttt{centroid\_com}) applied to the stellar mass maps, though we also find qualitatively similar results using a 2D Gaussian centroid estimator (\texttt{photuils.centroids} \texttt{centroid\_2dg}), or by simply using the galaxy Right Ascension and Declination given in the NASA/IPAC Extragalactic Database\footnote{\url{http://ned.ipac.caltech.edu/}}.

\subsection{Measuring asymmetries} \label{sec:methods_asym}

\begin{figure}
    \centering
    \includegraphics[width=\columnwidth]{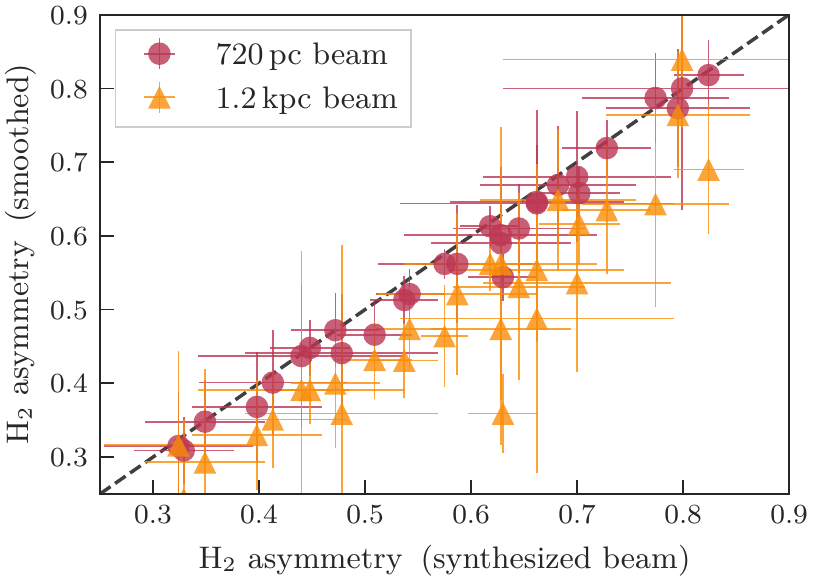}
    \caption{\htwo{} asymmetry measured from smoothed images at $720\text{-}\mathrm{pc}$ (red circles) and $1.2\text{-}\mathrm{kpc}$ (gold triangles) resolution versus \htwo{} asymmetry measured from images at the native, synthesized beam size. Dashed line shows the one-to-one line. With increasing beam size the measured asymmetry becomes smaller (i.e.\ falls further below the dashed line).}
    \label{fig:asym_beamsize}
\end{figure}

\begin{table*}[!th]
\footnotesize
\centering
\begin{threeparttable}
\caption{Galaxy asymmetry measurements}
\label{tab:galaxy_table}
\begin{tabular}{c c c c c c c c}
\toprule
\toprule
Galaxy & $\log_{10} (M_\bigstar / \mathrm{M_\odot})$ & $i$ & $A_\mathrm{mod}$ & $A_\mathrm{mod}$ & $A_\mathrm{mod}$ & $\theta_{\mathrm{low\text{-}res}}$ & Flags \\
& & ($^\circ$) & (\htwo{}, $720\,\mathrm{pc}$) & (\htwo{}, low-res) & (\hi{}, low-res) & (kpc) & \\
(1) & (2) & (3) & (4) & (5) & (6) & (7) & (8) \\
\midrule
Virgo Sample &&&&& \\
\midrule
IC3392 & 9.5 & 67.5 & $0.65 \pm 0.15$ & --- & --- & 1.44 & a,b \\
NGC4064 & 9.5 & 69.6 & $0.42 \pm 0.13$ & --- & --- & 1.34 & a,b \\
NGC4189 & 9.8 & 42.5 & $0.47 \pm 0.05$ & $0.39 \pm 0.09$ & $0.58 \pm 0.08$ & 1.35 & \\
NGC4254 & 10.5 & 38.8 & $0.61 \pm 0.03$ & $0.56 \pm 0.04$ & $0.69 \pm 0.04$ & 1.26 & \\
NGC4294 & 9.4 & 73.8 & $0.68 \pm 0.09$ & $0.52 \pm 0.12$ & $0.68 \pm 0.15$ & 1.34 & \\
NGC4298 & 10.1 & 52.3 & $0.35 \pm 0.06$ & $0.29 \pm 0.10$ & $0.45 \pm 0.09$ & 1.35 & \\
NGC4299 & 9.1 & 13.6 & $0.68 \pm 0.07$ & $0.55 \pm 0.12$ & $0.71 \pm 0.03$ & 1.34 & \\
NGC4321 & 10.7 & 32.2 & $0.45 \pm 0.04$ & $0.38 \pm 0.05$ & $0.37 \pm 0.03$ & 1.36 & \\
NGC4351 & 9.4 & 47.7 & $0.65 \pm 0.08$ & $0.56 \pm 0.11$ & $0.73 \pm 0.09$ & 1.34 & \\
NGC4380 & 10.1 & 60.6 & $0.32 \pm 0.07$ & $0.32 \pm 0.10$ & $0.49 \pm 0.08$ & 1.34 & \\
NGC4383 & 9.4 & 55.8 & $0.64 \pm 0.13$ & $0.48 \pm 0.14$ & $0.45 \pm 0.04$ & 1.26 & \\
NGC4394 & 10.3 & 31.9 & $0.72 \pm 0.04$ & $0.63 \pm 0.06$ & $0.30 \pm 0.07$ & 1.35 & \\
NGC4405 & 9.8 & 46.1 & $0.80 \pm 0.17$ & $0.82 \pm 0.13$ & --- & 1.35 & b \\
NGC4419 & 10.1 & 73.6 & $0.60 \pm 0.09$ & $0.53 \pm 0.12$ & $0.72 \pm 0.14$ & 1.36 & \\
NGC4424 & 9.9 & 61.5 & $0.85 \pm 0.06$ & --- & $0.77 \pm 0.11$ & 1.60 & a \\
NGC4450 & 10.7 & 51.1 & $0.66 \pm 0.04$ & $0.61 \pm 0.07$ & $0.42 \pm 0.07$ & 1.35 & \\
NGC4457 & 10.4 & 36.5 & $0.44 \pm 0.08$ & $0.36 \pm 0.13$ & $0.36 \pm 0.08$ & 1.43 & \\
NGC4501 & 11.0 & 65.5 & $0.40 \pm 0.07$ & $0.35 \pm 0.10$ & $0.60 \pm 0.08$ & 1.36 & \\
NGC4532 & 9.2 & 63.7 & $0.59 \pm 0.06$ & $0.44 \pm 0.10$ & $0.67 \pm 0.07$ & 1.44 & \\
NGC4535 & 10.5 & 48.4 & $0.51 \pm 0.03$ & $0.41 \pm 0.06$ & $0.48 \pm 0.03$ & 1.60 & \\
NGC4536 & 10.2 & 73.8 & $0.56 \pm 0.02$ & $0.44 \pm 0.05$ & $0.40 \pm 0.07$ & 1.52 & \\
NGC4548 & 10.7 & 36.8 & $0.54 \pm 0.03$ & $0.36 \pm 0.06$ & $0.49 \pm 0.08$ & 1.34 & \\
NGC4561 & 9.1 & 28.2 & $0.95 \pm 0.02$ & $0.92 \pm 0.08$ & $0.53 \pm 0.08$ & 1.27 & \\
NGC4569 & 10.9 & 68.7 & $0.67 \pm 0.08$ & $0.65 \pm 0.12$ & $0.74 \pm 0.11$ & 1.33 & \\
NGC4579 & 10.9 & 39.6 & $0.31 \pm 0.04$ & $0.22 \pm 0.07$ & $0.42 \pm 0.05$ & 1.59 & \\
NGC4580 & 9.9 & 46.1 & $0.35 \pm 0.13$ & --- & --- & 1.44 & a,b \\
NGC4606 & 9.6 & 68.6 & $0.82 \pm 0.05$ & $0.65 \pm 0.07$ & --- & 1.35 & b \\
NGC4651 & 10.3 & 52.9 & $0.56 \pm 0.08$ & $0.52 \pm 0.11$ & $0.32 \pm 0.06$ & 1.34 & \\
NGC4654 & 10.3 & 60.8 & $0.47 \pm 0.04$ & $0.43 \pm 0.08$ & $0.70 \pm 0.02$ & 1.35 & \\
NGC4689 & 10.2 & 38.5 & $0.37 \pm 0.07$ & $0.32 \pm 0.11$ & $0.28 \pm 0.09$ & 1.34 & \\
NGC4694 & 9.9 & 61.5 & $0.77 \pm 0.08$ & $0.76 \pm 0.12$ & $0.90 \pm 0.01$ & 1.34 & \\
NGC4698 & 10.5 & 66.0 & $0.61 \pm 0.06$ & $0.53 \pm 0.06$ & $0.38 \pm 0.05$ & 1.36 & \\
NGC4713 & 9.3 & 44.6 & $0.52 \pm 0.03$ & $0.41 \pm 0.07$ & $0.48 \pm 0.02$ & 2.08 & \\
NGC4772 & 10.2 & 59.8 & $0.79 \pm 0.06$ & $0.62 \pm 0.11$ & $0.50 \pm 0.04$ & 1.43 & \\
NGC4808 & 9.6 & 71.8 & $0.44 \pm 0.10$ & $0.38 \pm 0.11$ & $0.61 \pm 0.02$ & 1.36 & \\
\midrule
Control Sample &&&&&& \\
\midrule
NGC0628 & 10.2 & 29.5 & $0.55 \pm 0.01$ & $0.50 \pm 0.01$ & $0.45 \pm 0.03$ & 1.40 & \\
NGC2403 & 9.6 & 57.9 & $0.38 \pm 0.07$ & $0.30 \pm 0.09$ & $0.25 \pm 0.06$ & 1.40 & \\
NGC2841 & 10.9 & 65.5 & --- & $0.62 \pm 0.07$ & $0.40 \pm 0.02$ & 1.40 & c \\
NGC2903 & 10.4 & 61.9 & $0.41 \pm 0.09$ & $0.43 \pm 0.09$ & $0.36 \pm 0.05$ & 1.40 & \\
NGC3184 & 10.4 & 13.0 & --- & $0.48 \pm 0.03$ & $0.33 \pm 0.04$ & 1.40 & c \\
NGC3198 & 10.1 & 74.4 & --- & $0.39 \pm 0.07$ & $0.34 \pm 0.03$ & 1.40 & c \\
NGC3351 & 10.3 & 41.3 & $0.59 \pm 0.06$ & $0.47 \pm 0.03$ & $0.36 \pm 0.05$ & 1.40 & \\
NGC3521 & 10.8 & 53.8 & --- & $0.45 \pm 0.04$ & $0.39 \pm 0.05$ & 1.40 & c \\
NGC3627 & 10.7 & 59.1 & $0.53 \pm 0.04$ & $0.42 \pm 0.04$ & $0.46 \pm 0.09$ & 1.40 & \\
NGC3938 & 10.3 & 30.9 & --- & $0.41 \pm 0.02$ & --- & 1.40 & c,d \\
NGC4214 & 8.6 & 38.0 & $0.42 \pm 0.12$ & --- & $0.59 \pm 0.09$ & 1.40 & a \\
NGC4725 & 10.8 & 61.0 & --- & $0.68 \pm 0.03$ & --- & 1.40 & c,d \\
NGC4736 & 10.3 & 45.0 & $0.41 \pm 0.07$ & $0.51 \pm 0.04$ & $0.43 \pm 0.06$ & 1.40 & \\
NGC5055 & 10.7 & 62.8 & $0.42 \pm 0.03$ & $0.40 \pm 0.04$ & $0.45 \pm 0.01$ & 1.40 & \\
NGC5457 & 10.4 & 30.8 & $0.55 \pm 0.03$ & $0.46 \pm 0.04$ & $0.42 \pm 0.02$ & 1.40 & \\
NGC6946 & 10.5 & 35.8 & $0.46 \pm 0.01$ & $0.42 \pm 0.02$ & $0.40 \pm 0.01$ & 1.40 & \\
NGC7331 & 11.0 & 71.1 & --- & $0.53 \pm 0.04$ & $0.40 \pm 0.08$ & 1.40 & c \\
\bottomrule
\end{tabular}
\begin{tablenotes}[flushleft]
    \item \textbf{Notes:} Table columns are: (1) galaxy identifier, (2) integrated stellar mass from \citet{leroy2019}, (3) inclination, (4) $A_\mathrm{mod}$ from $720\text{-}\mathrm{pc}$ \htwo{} map, (5) $A_\mathrm{mod}$ from low-res \htwo{} map, (6) $A_\mathrm{mod}$ from low-res \hi{} map, (7) low-res FWHM beam size, (8) flags explaining missing $A_\mathrm{mod}$ values.
    \\[-0.5em]
    \item[a] <5 beams across low-res CO map
    \item[b] <5 beams across low-res \hi{} map
    \item[c] No $720\,\mathrm{pc}$ CO image
    \item[d] Not in THINGS sample
\end{tablenotes}
\end{threeparttable}
\end{table*}

A standard quantitative measure of galaxy asymmetry is the rotational asymmetry, with the asymmetry parameter, $A$, given by:
\begin{equation} \label{eq:asym_classic}
    A = \frac{\sum_{i,j}|I(i,j) - I_{180}(i,j) |}{\sum_{i,j}|I(i,j)|}
\end{equation}
\noindent
where $I(i,j)$ is the original image, $I_{180}(i,j)$ is the image after a $180^\circ$ rotation, and the sums are over all pixel indices $i,j$. Thus $A$ measures the fraction of the total flux that is contained in asymmetric components. This technique was first applied to rest-frame optical images of galaxies \citep[e.g.][]{schade1995,abraham1996,conselice2003_cas}, but has since been applied to \hi{} images of galaxies to quantify atomic gas asymmetries \citep[e.g.][]{holwerda2011_things,lelli2014,giese2016,reynolds2020,bilimogga2022} and recently also to molecular gas images \citep{davis2022,lee2022}.  Because the classic asymmetry parameter, $A$, is normalized by the total flux, brighter pixels that are typically close to the galaxy centre contribute more weight to the asymmetry parameter than faint pixels in the galaxy outskirts. In order to obtain a more uniform weighting between pixels near the galaxy centre and pixels on the outskirts, \citet{lelli2014} introduced a modified asymmetry parameter ($A_\mathrm{mod}$, see also \citealt{bilimogga2022, deb2023}) that is given by:
\begin{equation} \label{eq:Amod}
    A_\mathrm{mod} = \frac{1}{N} \sum\limits_{i,j}^{N} |\delta_{180}|,
\end{equation}
where
\begin{equation}
    \delta_{180} = \frac{I(i,j) - I_{180}(i,j)}{I(i,j) + I_{180}(i,j)},
\end{equation}
\noindent
$N$ is the total number of pixels in the image, and $\delta_{180}$ is the local difference map after a $180^\circ$ rotation. In this way, $A_\mathrm{mod}$ is normalized by the local flux as opposed to the total flux and is more sensitive to faint asymmetric features. As outlined in Section~\ref{sec:data_mstar}, we measure the rotational asymmetry about the stellar mass centroid for each galaxy. For this study we opt to use $A_\mathrm{mod}$ to quantify the asymmetry of \hi{} and \htwo{} gas surface density maps in order to ensure that we are sensitive to low-surface-brightness gas tails that can be observed behind cluster galaxies \citep[e.g.][]{kenney2004,oosterloo2005,chung2009,yoon2017,kenney2015,hess2022}.
\par
When calculating $A_\mathrm{mod}$ (for both Virgo and control galaxies), we employ a resolution criteria requiring that all galaxies have detected emission in their gas surface density maps that is at least five beam widths (FWHM) across. We note that for galaxies with only $\sim\! 5-10$ beams across, our measured value of $A_\mathrm{mod}$ may be systematically lower than the intrinsic value \citep{bilimogga2022}. We also require that galaxies have inclinations of $i < 75^\circ$ in order to exclude highly-inclined galaxies. Previous work has shown that model galaxies viewed at large inclinations tend to have small measured rotational asymmetries compared to the intrinsic value (i.e.\ face-on, e.g.\ \citealt{giese2016}). This inclination cut is aimed at striking a balance between reducing the biases in our measured asymmetries while still maintaining a relatively large sample. We have tested a stricter cut of $i < 60^\circ$ as well as the looser approach of no inclination cut at all. In either case, the qualitative conclusions from this work are insensitive to these different inclination restrictions. We opt for the intermediate case requiring $i < 75^\circ$, but do note that we may be underestimating the intrinsic asymmetry for galaxies near this limit.  Our final samples of Virgo galaxies and control galaxies used for the $A_\mathrm{mod}$ analysis are listed in Table \ref{tab:galaxy_table}.
\par
In Fig.~\ref{fig:gas_density_dist} we show distributions of $\mathrm{H_2}$ and \hi{} surface density consisting of all of the pixels that make up the gas surface density maps for the Virgo and control galaxies used for the asymmetry analysis. Overall the distributions for Virgo and control galaxies cover similar ranges in surface densities, both for $\mathrm{H_2}$ and \hi{}, with only small offsets in the medians of $0.05-0.1\,\mathrm{dex}$. Fig.~\ref{fig:gas_density_dist} is meant to convey the broad range of gas surface densities covered by the various samples from this work, though we note that observations of individual galaxies have varying depths even within each survey.  The specific sensitivities reached for each galaxy can be found in the overview papers for each survey \citep{walter2008,chung2009,leroy2009,brown2021}. There is a small tail in the VERTICO distribution at small surface densities ($\log_{10}[\Sigma_\mathrm{H_2} / \mathrm{M_\odot\,pc^{-2}}] \lesssim 0.5)$ that is not replicated in the HERACLES distribution. To ensure that this small difference between the VERTICO and HERACLES gas surface density distributions does not bias our results, we re-calculate the \htwo{} asymmetry for VERTICO galaxies but only including pixels with $\log_{10}[\Sigma_\mathrm{H_2} / \mathrm{M_\odot\,pc^{-2}}] > 0.5$. In doing this we find a minimal effect on the measured asymmetries and that the qualitative results (e.g. from Fig.~\ref{fig:asym_HIclass}) are not changed.
\par
We consider two sources of error when calculating uncertainties on $A_\mathrm{mod}$ for each galaxy. First, a statistical error on the pixel-by-pixel flux measurements, and second an error related to the choice of central pixel for the calculation of $A_\mathrm{mod}$. For the flux error, $\sigma_\mathrm{flux}$, we apply random bootstrap re-sampling to the computation of $A_\mathrm{mod}$ from equation~\ref{eq:Amod}. We then take $\sigma_\mathrm{flux}$ to be the standard deviation of the resulting bootstrap distribution. We measure $A_\mathrm{mod}$ about the stellar mass centroid for each galaxy, but in our uncertainty analysis we also consider the effect of shifting central pixel used. For each galaxy we define a set of `central pixels' which correspond to all pixels within a circle centred on the stellar mass centroid with a diameter equal to the FWHM beam size. This FWHM corresponds to the specific beam size of the map that the asymmetry was measured from (see Section~\ref{sec:data_resolution}). We then re-compute $A_\mathrm{mod}$ but now rotating the image about each of these central pixels, giving a list of `shifted' $A_\mathrm{mod}$ values. Our estimate for the centroiding contribution to the uncertainty on $A_\mathrm{mod}$, $\sigma_\mathrm{centroid}$, is then given by the standard deviation of this set of shifted $A_\mathrm{mod}$ values. Finally, the total error on $A_\mathrm{mod}$ that we quote for each galaxy is given by the sum of $\sigma_\mathrm{flux}$ and $\sigma_\mathrm{centroid}$ in quadrature.

In Fig.~\ref{fig:example_img} we show optical, \htwo{}, and \hi{} images for an example galaxy from our sample, NGC4654. As a visual reference for $A_\mathrm{mod}$, we also show maps of $\delta_{180}$ as applied to the CO and \hi{} gas surface density maps. For \hi{}, the one-sided tail to the southeast stands out in the $\delta_{180}$ map as a large group of pixels with $\delta_{180} = +1.0$. There is also a relatively bright gas feature on the northwest edge of NGC4654 that is visible in the $\delta_{180}$ map for both \hi{} and CO. Table~\ref{tab:galaxy_table} lists our measured values for $A_\mathrm{mod}$ for all Virgo and control galaxies that satisfy the aforementioned quality cuts.

\subsubsection{A note on resolution} \label{sec:data_resolution}

Across a range of wavelengths it has been shown that lower resolutions lead to artificially small values for measurements of rotational asymmetries \citep[e.g.][]{bershady2000,conselice2000,giese2016,thorp2021} and this issue still persists for $A_\mathrm{mod}$ \citep{bilimogga2022}. With this in mind, in this work we always measure asymmetries at the highest resolution possible, while still ensuring that a fair comparison between \htwo{} and \hi{} asymmetries are possible where necessary. This ensures that we are using \htwo{} asymmetries that are as close as possible to the intrinsic \htwo{} asymmetry for the galaxies in our sample, as again, degrading resolution leads to a systematic underestimate of intrinsic asymmetry. We demonstrate this effect directly for our sample in Fig.~\ref{fig:asym_beamsize}, where we plot the asymmetry measured at $720\text{-}\mathrm{pc}$ and $1.2\text{-}\mathrm{kpc}$ resolution as a function of the asymmetry measured at the native resolution of the VERTICO maps (see Table 2 in \citealt{brown2021}). This figure demonstrates the decreasing measured asymmetry with increasing beam size, though there is little difference between the native resolution and $720\text{-}\mathrm{pc}$ resolution asymmetries as the native beam sizes for VERTICO galaxies are only slightly smaller than $720\,\mathrm{pc}$.
\par
As is outlined in Section~\ref{sec:data_viva}, the VIVA \hi{} products used in this work have physical resolutions ranging between $1.2\,\mathrm{kpc}$ and $2.0\,\mathrm{kpc}$, with an average resolution of $1.4\,\mathrm{kpc}$ (see Table~\ref{tab:galaxy_table} for the specific beam size for each galaxy).  Thus for this work we define two resolution-based samples for Virgo galaxies:  a $720\text{-}\mathrm{pc}$ resolution \htwo{}-only sample, and a low-resolution (``low-res'') \htwo{}$+$\hi{} sample where the \htwo{} maps are smoothed to match the physical resolution of the VIVA \hi{} products on a galaxy-by-galaxy basis. This low-res sample is used when making explicit, quantitative comparisons between \htwo{} and \hi{} asymmetries.
\par
As mentioned in Section~\ref{sec:data_heracles}, we also construct a $720\text{-}\mathrm{pc}$ resolution \htwo{}-only control sample from HERACLES and a ``low-res'' \htwo{}$+$\hi{} control sample where we smooth the corresponding HERACLES and THINGS maps to a physical resolution of $1.4\,\mathrm{kpc}$. This corresponds to the average physical resolution of the low-res Virgo sample.
\par
Lastly, SFR and stellar mass surface density maps are produced at both $720\text{-}\mathrm{pc}$ and $1.4\text{-}\mathrm{kpc}$ resolution in order to match the \htwo{} and \hi{} products.  For every figure in the paper we make clear in the caption which resolution sample is being used.

\section{Virgo galaxy cold gas asymmetries} \label{sec:virgo_asym}

\subsection{Asymmetry versus \textsc{Hi} stripping class} \label{sec:asym_HIclass}

\begin{figure*}
    \centering
    \includegraphics[width=0.48\textwidth]{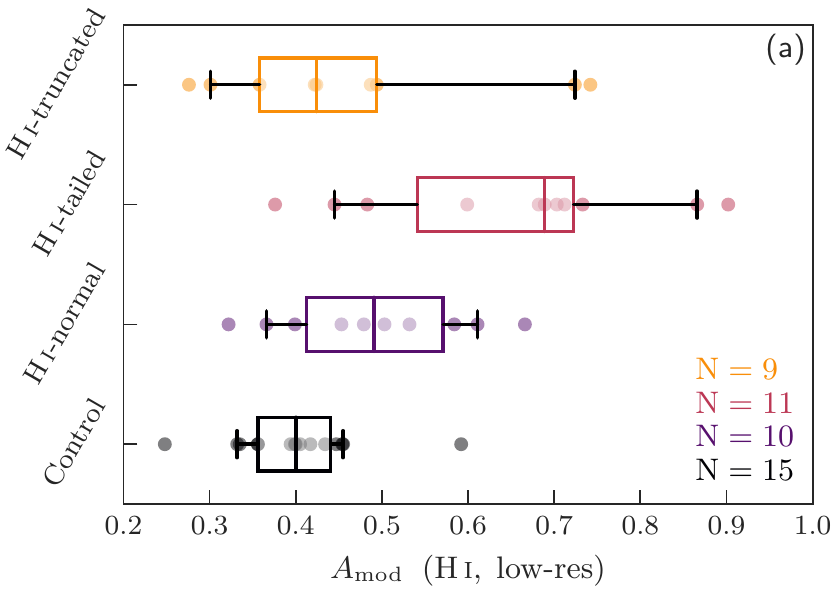}
    \hfill
    \includegraphics[width=0.48\textwidth]{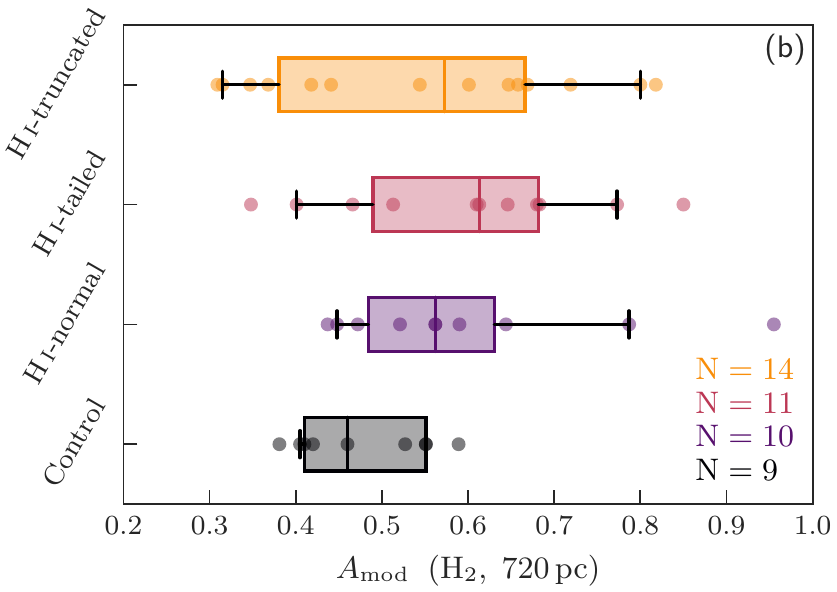}
    \caption{Gas asymmetries as a function of \citet{yoon2017} \hi{} stripping class. Left-hand panel shows individual measurements (circles) as well as box plots for atomic gas asymmetries from VIVA and THINGS (control) \hi{} maps from the low-res sample (average resolution, $1.4\,\mathrm{kpc}$). Right-hand panel shows individual measurements as well as box plots for molecular gas asymmetries measured from VERTICO and HERACLES (control) \htwo{} maps at $720\text{-}\mathrm{pc}$ resolution. For the box plots: boxes span the interquartile range for each class, the solid vertical line marks the median, the whiskers range from the 5th to the 95th percentile of the distribution. The numbers in the lower right of each panel show the number of galaxies in each class that pass the requisite quality cuts (see Section~\ref{sec:methods_asym}). While there are similar qualitative patterns between \hi{} and \htwo{} asymmetries, the magnitude of the difference between classes is clearly larger for \hi{} asymmetry.}
    \label{fig:asym_HIclass}
\end{figure*}

We test whether the quantitative \hi{} and \htwo{} asymmetries systematically differ for galaxies in different \hi{} morphological classes, using a modified version of the \hi{} classes for VIVA galaxies that were defined by \citet{yoon2017}. The \citet{yoon2017} classes use morphological classifications in an attempt to separate different stages of atomic gas stripping. These classes include \hi{}-normal galaxies with symmetric atomic gas distributions (``Class 0''), galaxies with atomic gas tails that are relatively \hi{}-normal (``Class I'') or relatively \hi{}-deficient (``Class II''), galaxies with symmetric atomic gas distributions that are severely truncated within the stellar disc (``Class III''), and galaxies with symmetric atomic gas distributions that are marginally truncated within the stellar disc (``Class IV''). In this work we merge some of these qualitative \hi{} classes from \citet{yoon2017} in order to increase the number of galaxies per class, resulting in three qualitative \hi{} morphology classes that we will use moving forward: ``\hi{}-normal'' (\citeauthor{yoon2017} Class 0), ``\hi{}-tailed'' (\citeauthor{yoon2017} Classes I and II), ``\hi{}-truncated'' (\citeauthor{yoon2017} Classes III and IV). We note that these are identical to the ``Unperturbed'', ``Asymmetric'', and ``Symmetric-truncated'' \hi{} classes used in \citet{villanueva2022}. These classes can be understood to roughly trace three different stages of environmental influence, with \hi{}-normal tracing galaxies that appear largely unperturbed by environment, \hi{}-tailed tracing galaxies that are near peak environmental perturbation, and \hi{}-truncated tracing galaxies that have already experienced substantial environmental influence. We note that our \hi{}-truncated class includes both galaxies with severely truncated \hi{} discs, likely a product of RPS (Class III), as well as galaxies with more marginal truncation that are consistent with gradual gas depletion associated with starvation (Class IV). 
\par
In Fig.~\ref{fig:asym_HIclass}a we show box plots of \hi{} asymmetry for galaxies split into the non-Virgo control sample as well as the three \hi{} morphological classes. The Virgo \hi{} asymmetries are measured from gas surface density maps with physical resolutions between $1.2\,\mathrm{kpc}$ and $2.0\,\mathrm{kpc}$ (with an average of $1.4\,\mathrm{kpc}$), and the control \hi{} asymmetries are measured at a common resolution of $1.4\,\mathrm{kpc}$.  Galaxies in the control sample still show significant asymmetries in their \hi{} distributions, with a median of $A_\mathrm{mod} \simeq 0.4$ and a distribution which is clearly greater than zero. This is consistent with previous works showing that even `typical' galaxies have measurable asymmetries in their \hi{} distributions \citep{richter1994,haynes1998,matthews1998,watts2020a,watts2020b}. Relative to environment, however, the distribution of \hi{} asymmetry for the control sample is centred on a smaller value than the distributions for the various Virgo galaxy classes. On average, Virgo galaxies have \hi{} asymmetries that are larger than the control sample by $40 \pm 10$ per cent. We also apply a two-sample Anderson-Darling test \citep{anderson1952,scholz1987} to the \hi{} asymmetry measurements for the Virgo and control samples and find evidence at the $99.8$ per cent significance level that the two distributions are distinct.
\par
Relative to the control, galaxies in the \hi{}-normal class show enhanced asymmetries, hinting that there is still some impact from environment on \hi{} distributions even at this early phase of environmental evolution. Compared to the other classes, the \hi{} asymmetry distribution for \hi{}-tailed galaxies is clearly skewed to the largest values. Of course, this is unsurprising given that these galaxies are selected on the basis of one-sided \hi{} extensions which in turn drive large values of $A_\mathrm{mod}$. Galaxies in the \hi{}-truncated class have median asymmetries that are comparable to the control sample, this reflects their symmetric but compact \hi{} morphologies.  There are two \hi{}-truncated galaxies that still show large asymmetries ($A_\mathrm{mod} > 0.7$, NGC4419 and NGC4569).  In both cases these are compact \hi{} distributions with asymmetries about the galaxy centre well within the galaxy disc. While the differences in Fig. \ref{fig:asym_HIclass}a are in part by construction (due to the morphological nature of our modified \citealt{yoon2017} classes), we explore a more interesting question in Fig. \ref{fig:asym_HIclass}b. Namely, whether or not the asymmetry of molecular gas in Virgo galaxies differ as a function of atomic-gas-stripping stage.
\par
Gas-stripping phenomena in clusters, such as RPS, should more strongly perturb atomic gas reservoirs in galaxies compared to molecular gas. \hi{} has a relatively extended distribution in galaxies making it susceptible to stripping, whereas molecular gas is less efficiently stripped for two main reasons: first, it is primarily located in giant molecular clouds that have a smaller cross-section to the external pressure and are much denser than the diffuse \hi{}, and second, molecular gas is concentrated within the inner disc where the gravitational potential well of the galaxy is deeper \citep[e.g.][]{solomon1987,engargiola2003,bolatto2008,yamagami2011,cortese2021,boselli2022_review}. Thus it is expected that the molecular gas component of galaxies will be less impacted by environmental effects compared to the atomic component. Indeed, this picture is consistent with other early results from the VERTICO survey \citep[e.g.][]{zabel2022,watts2023}.
\par
In Fig. \ref{fig:asym_HIclass}b we plot the distributions of the measured \htwo{} asymmetry for the control sample and the three \hi{} stripping classes. These \htwo{} asymmetries are measured at a physical resolution of $720\,\mathrm{pc}$ ($9\arcsec$ for Virgo galaxies).  The variations in asymmetry between different classes for \htwo{} in Fig.~\ref{fig:asym_HIclass}b are less apparent than for \hi{} asymmetry in Fig.~\ref{fig:asym_HIclass}a. The \htwo{} asymmetry distributions for the control sample and for the Virgo galaxies show more overlap than for \hi{}, thus we measure a smaller environmental imprint on molecular gas distributions relative to atomic. Statistically, there is still weak evidence ($\sim\!2\sigma$) for a difference between \htwo{} asymmetries for Virgo galaxies compared the control, with the average value of $A_\mathrm{mod}$ for Virgo galaxies being $20 \pm 10$ per cent larger than for the control sample. Additionally, a two-sample Anderson-Darling test \citep{anderson1952,scholz1987} finds evidence at the 96 per cent confidence level that the distributions of \htwo{} asymmetry for the Virgo and control samples were drawn from distinct parent distributions. Focusing in on Virgo galaxies, there are no strong differences between the \htwo{} asymmetries for galaxies in different \hi{} morphological classes, with all three distributions showing substantial overlap. The outlier \hi{}-normal galaxy with $A_\mathrm{mod} > 0.9$ is NGC4561, an interesting galaxy with a large, extended \hi{} distribution in VIVA but only a few knots of CO emission detected sporadically across the galaxy disc in VERTICO.  
\par 
The results in Fig.~\ref{fig:asym_HIclass}a are in agreement with \citet{reynolds2020} who use \hi{} observations from the Local Volume \hi{} Survey (LVHIS, \citealt{koribalski2018}), the Hydrogen Accretion in Local Galaxies Survey (HALOGAS, \citealt{heald2011}), and VIVA in order to explore variations in various measures of \hi{} asymmetry as a function of environmental density. Most relevant to this work is the fact that \citeauthor{reynolds2020} show that their $A_\mathrm{map}$ asymmetry parameter (analogous to the classical asymmetry parameter given by equation~\ref{eq:asym_classic}) is systematically enhanced for VIVA galaxies (i.e.\ the cluster environment) in comparison to galaxies from LVHIS and HALOGAS that are more representative of galaxy groups and isolated galaxy environments.  \citet{reynolds2020} do not consider $A_\mathrm{mod}$ among their asymmetry parameters, so we cannot make a direct comparison with this work, but the qualitative trends are in agreement. We are also broadly consistent with work measuring asymmetries of \hi{} spectra from the xGASS observational survey \citep{catinella2010,catinella2018} as well as the IllustrisTNG simulation \citep[e.g.][]{pillepich2018,nelson2019}, that find enhanced \hi{} spectral asymmetries for satellite galaxies in overdense environments \citep{watts2020a,watts2020b}.
\par
While this work adds to the literature consensus that environmental effects impart measurable \hi{} asymmetries in satellite galaxies, the more pressing question that we aimed to address in this section is whether or not asymmetries of the denser, more centrally concentrated molecular gas component show systematic trends with environmental influence. For \htwo{} asymmetries in Fig.~\ref{fig:asym_HIclass}b, the picture is less clear. We do find some evidence that \htwo{} asymmetries for Virgo galaxies are slightly larger than for the control sample, but this is only marginally significant. Furthermore, there is no clear trend of CO asymmetry with the degree of environmental influence, as traced by the \citet{yoon2017} \hi{} morphological classes. In particular, \hi{}-normal and \hi{}-tailed galaxies show very similar \htwo{} asymmetry distributions.  This may point to environmental perturbations on \htwo{} only becoming strongly apparent after substantial \hi{} stripping.  This is consistent with the collection of \hi{}-truncated galaxies with large \htwo{} asymmetries in Fig.~\ref{fig:asym_HIclass}b.

\subsection{Direct comparison between \hi{} and CO asymmetries} \label{sec:asym_hi_co}

\begin{figure}
    \centering
    \includegraphics[width=\columnwidth]{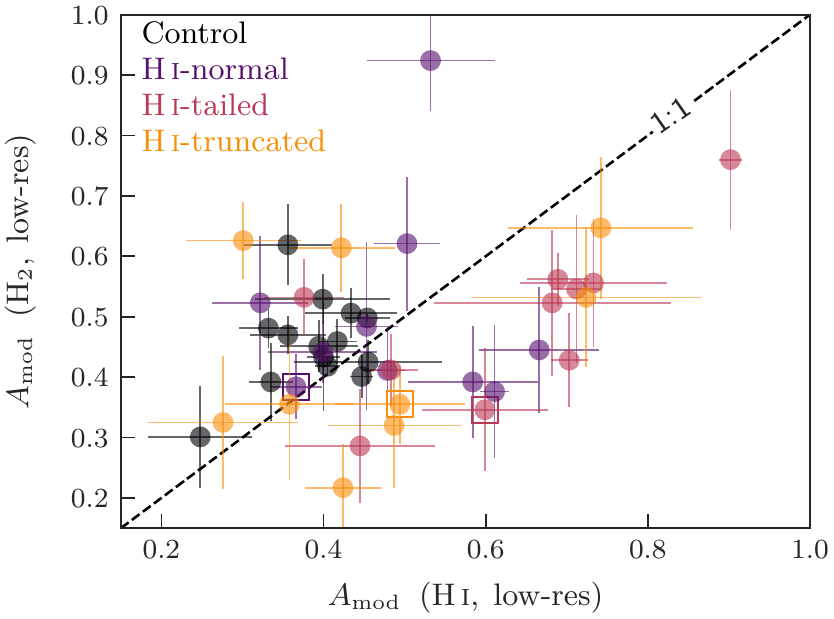}
    \caption{\htwo{} versus \hi{} asymmetry on a galaxy-by-galaxy basis at matched resolution. For Virgo galaxies, asymmetries are measured at the low-res beam sizes listed in Table~\ref{tab:galaxy_table}, for control galaxies asymmetries are measured at a common resolution of $1.4\,\mathrm{kpc}$. Galaxies in different \hi{} classes are denoted by different marker colours.  The dashed line corresponds to $A\mathrm{_{mod}(H_2)} = A\mathrm{_{mod}(H\,\textsc{i})}$.}
    \label{fig:asym_hi_co}
\end{figure}

We extend the comparison of \hi{}-classes in Section~\ref{sec:asym_HIclass} by directly comparing \hi{} and \htwo{} asymmetries on a galaxy-by-galaxy basis. When making direct, quantitative comparisons between \hi{} and \htwo{} asymmetries, it is critical that this comparison be made at matched resolution. We only include galaxies that have both robust \hi{} asymmetries and robust \htwo{} asymmetries from the low-res sample, which amounts to 29 galaxies from the Virgo galaxy sample and 14 galaxies from the control sample (see Table~\ref{tab:galaxy_table}).
\par
In Fig.~\ref{fig:asym_hi_co} we plot \htwo{} asymmetry versus \hi{} asymmetry with individual datapoints coloured according to the control sample and the three Virgo \hi{} classes. In Fig.~\ref{fig:asym_hi_co} we find that there is a weak correlation (Spearman correlation coefficient of $0.3$, $p\text{-value} = 0.06$) between \hi{} and \htwo{} asymmetry when considering all galaxies in our sample (both Virgo and control), with significant scatter. Again, the outlier toward the top of the panel with an \htwo{} asymmetry of $\sim\!0.9$ is NGC4561 (see Section~\ref{sec:asym_HIclass}).
\par
The control sample galaxies have an average \hi{}-to-\htwo{} asymmetry ratio ($A_\mathrm{mod,\,\hi{}} / A_\mathrm{mod,\,\htwo{}}$) of $0.86 \pm 0.04$. This may be a product of CO tracing a clumpier gas distribution than \hi{}. Though we stress that this is only a modest deviation from the one-to-one line and a limited number of galaxies. \hi{}-normal and \hi{}-truncated galaxies tend to scatter around both sides of the one-to-one line with average \hi{}-to-\htwo{} asymmetry ratios of $1.06 \pm 0.11$ and $1.16 \pm 0.14$. Finally, galaxies in the \hi{}-tailed class are more skewed toward large \hi{} asymmetries, with an average \hi{}-to-\htwo{} asymmetry ratios of $1.31 \pm 0.09$. All but one (NGC4698) of the \hi{}-tailed galaxies have larger \hi{} asymmetries than for \htwo{}.
\par
By eye, there are arguably two groups of points in Fig.~\ref{fig:asym_hi_co}, a cloud of points around the one-to-one line, centred roughly on $A_\mathrm{mod}\mathrm{(\hi{})} \simeq 0.4$ (with `normal' asymmetries) and a sequence of points below the one-to-one line forming a correlation at $A_\mathrm{mod}\mathrm{(H\,\textsc{i})} \gtrsim 0.5$. More quantitatively, if we consider galaxies with large \hi{} asymmetries (here we define `large' to mean above the median value of $A_\mathrm{mod}\mathrm{(H\,\textsc{i}}) = 0.45$), there is a moderately significant correlation between \hi{} and \htwo{} asymmetry according to the Spearman rank test ($r_s = 0.49$, $p\text{-value} = 0.02$).  Conversely, for galaxies with `small' \hi{} asymmetries (i.e.\ below the median value), there is no evidence for a correlation between \hi{} and \htwo{} asymmetry ($r_s = 0.12$, $p\text{-value} = 0.60$).
\par
This sequence with large asymmetries primarily consists of \hi{}-tailed galaxies though also contains a few galaxies from each of the \hi{}-normal (NGC4189, NGC4532, NGC4808) and \hi{} truncated (NGC4419, NGC4569) classes. These galaxies have \hi{} and \htwo{} asymmetries that are correlated, but at the same time always have larger \hi{} asymmetries than \htwo{}.  A natural interpretation for this is that these are galaxies experiencing strong environmental perturbations, but that have not yet been severely stripped of their \hi{} content. If true, then this is evidence for environmental perturbations to not only \hi{} (which is most strongly perturbed) but also \htwo{} given the correlation between \hi{} and \htwo{} asymmetries. We do stress that this particular analysis is subject to small-number statistics given our sample size, which is then halved to isolate the large-asymmetry galaxies. This interpretation should be read as tentative, bearing in mind that future works with larger samples will be able to confirm or rule out the robustness of this trend. 
\par
Overall, the results presented in Section~\ref{sec:virgo_asym} confirm that the signatures of environmental perturbations, which we traced with $A_\mathrm{mod}$, are more apparent when considering \hi{} compared to the denser molecular gas. We do find marginal evidence for larger \htwo{} asymmetries for Virgo galaxies compared to a non-cluster control sample, and also a correlation between \hi{} and \htwo{} asymmetry for the most \hi{} asymmetric galaxies in our sample. We interpret this as modest evidence that environmental perturbations are also imprinted on \htwo{} asymmetries of cluster galaxies, but a larger sample is necessary before this can be stated with confidence.

\section{CO emission on the leading and trailing Sides} \label{sec:mol_dense_lt}

\begin{figure}[!ht]
    \centering
    \includegraphics[width=\columnwidth]{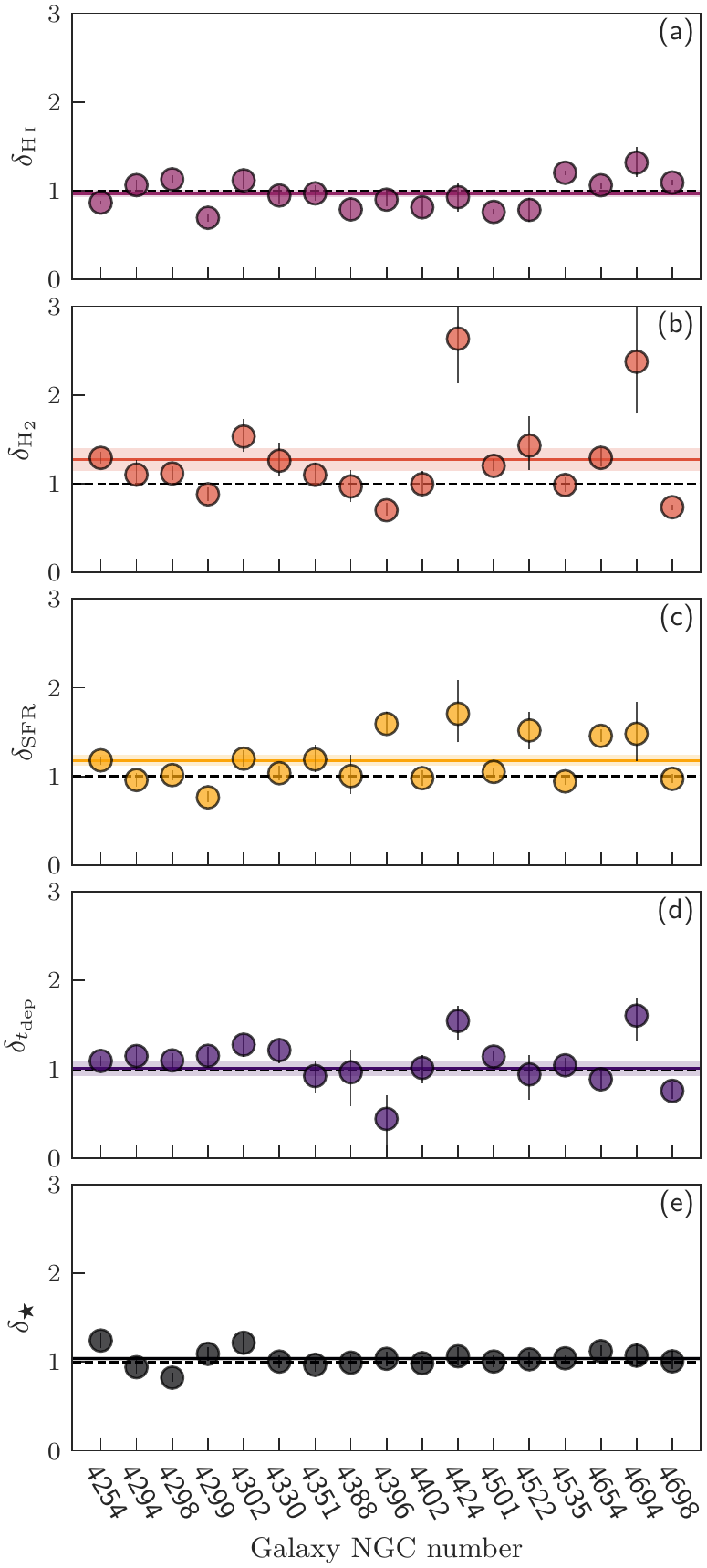}
    \caption{Anisotropies between the leading and trailing halves of \hi{}-tailed galaxies (see text for details). From top to bottom: atomic gas mass, molecular gas mass, star formation rate, molecular gas depletion time, and stellar mass. Solid lines and shaded bands mark the median and statistical error on the median. Both errors on the median and errorbars on the data points are derived from bootstrap resampling. Both \htwo{} and SFR show excess emission on the leading half.}
    \label{fig:lt_ratio}
\end{figure}

\begin{figure}
    \centering
    \includegraphics[width=\columnwidth]{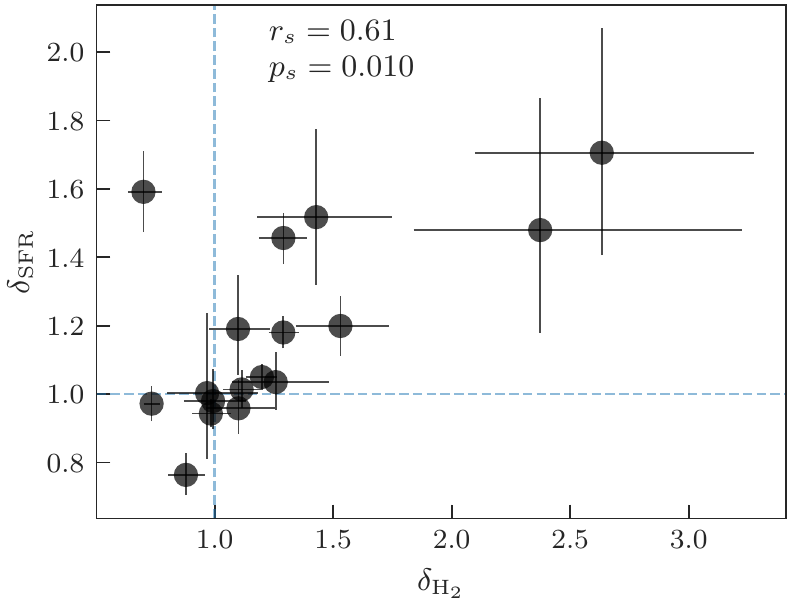}
    \caption{The SFR leading half anisotropy versus that same for molecular gas mass. The clear correlation highlights that galaxies with large leading half excesses of molecular gas also show similar excesses in star formation. The one notable outlier to the trend is NGC4396.}
    \label{fig:delta_h2_sfr}
\end{figure}

Previous studies, both from an observational and theoretical perspective, have argued that ram pressure is capable of compressing existing molecular gas as well as promoting the conversion from atomic to molecular gas in cluster galaxies \citep[e.g.][]{bekki2014,cramer2020,moretti2020b,troncoso-iribarren2020}. In principle, signatures of this ISM compression should be strongest on the `leading half' of galaxies experiencing strong ram pressure. In simulations the leading and trailing halves can be separated according to the orbital velocity vector between the galaxy and the ICM. But observationally, where full phase-space information cannot be obtained, the best proxy for the direction of motion for a galaxy is the orientation of a stripped tail. In this way cluster galaxies with observed tails can be roughly divided into a leading and trailing half, modulo projection effects (see e.g.\ \citealt{troncoso-iribarren2020,lee2022_virgo_tails,roberts2022_lofar_manga}, for other examples of this sort of division). We note that this technique will be most sensitive to perturbations occurring recently relative to the rotational period of a galaxy, as over longer timescales gas which was compressed on the leading half could be transported to trailing half (if not already consumed via star formation).
\par
The VERTICO survey contains 18 galaxies with \hi{} tails (i.e.\ Class I or Class II from \citealt{yoon2017}), and therefore we use this sample to test these predictions of increased ISM densities on the leading versus the trailing half. For this section we include all \hi{}-tailed galaxies in VERTICO and do not employ the inclination restriction ($i < 75^\circ$) that was used in Section~\ref{sec:virgo_asym} to avoid biasing measured values of $A_\mathrm{mod}$. We note that a clear majority of the \hi{}-tailed galaxies used here are believed to be undergoing RPS \citep{chung2007,yoon2017,boselli2022_review}, though we cannot rule out that gravitational interactions are also playing a role in the gas stripping for some of these galaxies. In order to split the \hi{}-tailed galaxies into a leading and trailing half we use the divisions published by \citet{lee2022_virgo_tails} that are based on observed \hi{} tail directions (see \citealt{lee2022_virgo_tails} for details). The dividing line between the leading and trailing half is then taken to be the line that passes through the galaxy centre, which in our case is the stellar mass centroid, and is normal to the \hi{} tail direction (see Fig.~\ref{fig:example_img} for an illustration of this division).
\par
We consider a quantitative measure of the difference in \htwo{} mass between the leading and trailing halves,
\begin{equation}
    \delta_\mathrm{H_2} \equiv \frac{M_\mathrm{{H_2,\,LH}}}{M_\mathrm{{H_2,\,TH}}},
\end{equation}
\noindent where $M_\mathrm{H_2,\,LH}$ and $M_\mathrm{H_2,\,TH}$ are the total \htwo{} gas masses on the leading and trailing halves, respectively. We also define four analogous quantities in terms of \hi{} mass, SFR, molecular gas depletion time ($t_\mathrm{dep}$), and stellar mass given by:
\begin{equation}
    \delta_\mathrm{H\,\textsc{i}} \equiv \frac{M_\mathrm{H\,\textsc{i},\,LH}}{M_\mathrm{H\,\textsc{i},\,TH}},
\end{equation}
\begin{equation}
    \delta_\mathrm{SFR} \equiv \frac{\mathrm{SFR}_\mathrm{LH}}{\mathrm{SFR}_\mathrm{TH}},
\end{equation}
\begin{equation}
    \delta_{t_\mathrm{dep}} \equiv \frac{M_\mathrm{{H_2,\,LH}}\,/\,\mathrm{SFR_\mathrm{LH}}}{M_\mathrm{{H_2,\,TH}}\,/\,\mathrm{SFR_\mathrm{TH}}},
\end{equation}
\noindent and
\begin{equation}
    \delta_{\bigstar} \equiv \log \frac{M_{\bigstar,\,\mathrm{LH}}}{M_{\bigstar,\,\mathrm{TH}}}.
\end{equation}
\noindent
When calculating these delta quantities, we only include pixels within the optical galaxy radius, taken to be $r_{25}$ from the Hyperleda database\footnote{https://leda.univ-lyon1.fr} \citep{makarov2014}. Thus this analysis focuses on anisotropies within the galaxy disc, which is where signatures from potential gas compression are expected to occur. In Fig.~\ref{fig:lt_ratio} we show $\delta_\mathrm{\hi{}}$ (a), $\delta_\mathrm{H_2}$ (b), $\delta_\mathrm{SFR}$ (c), $\delta_{t_\mathrm{dep}}$ (d), and $\delta_\bigstar$ (e) for each of the \hi{}-tailed galaxies.
\par
There is no systematic difference between the total atomic gas mass on the leading and trailing halves, in other words, $\delta_\mathrm{\hi{}}$ is centred on one. However, the fact that we consider total gas mass when calculating $\delta_\mathrm{\hi{}}$ does mask some differences between atomic gas properties on the leading and trailing halves. Namely, the average atomic gas surface density is systematically larger on the leading half relative to the trailing half, by a modest amount. Typically, by a factor of $\sim\!10$ per cent and up to $\sim\!50$ per cent for some galaxies. Conversely, these galaxies also have truncated \hi{} emission on their leading halves compared to their trailing halves (a product of gas stripping), and thus the combined effect leads to a total atomic gas mass which similar between the two halves.
\par
Unlike for \hi{}, there is modest evidence for excess molecular gas mass on the leading half as shown in Fig.~\ref{fig:lt_ratio}b. The distribution of $\delta_\mathrm{H_2}$ is skewed to values greater than one and the median $\delta_\mathrm{H_2}$ (solid line) is greater than one with a significance level between 2 and $3\sigma$. We note that this molecular gas excess on the leading half remains when considering the average molecular gas surface density on the leading and trailing halves instead of the total molecular gas mass as is shown in Fig.~\ref{fig:lt_ratio}b. The two galaxies in Fig.~\ref{fig:lt_ratio}b with $\delta_\mathrm{H_2} > 2$ are NGC4424 and NGC4694. In both cases the large $\delta$-value is driven by a single, large \htwo{}-clump slightly offset from the galaxy centre on the leading half. There is evidence that both NGC4424 and NGC4694 have experienced a recent mergers and/or tidal interactions in addition to ongoing RPS \citep[e.g.][]{cortes2006,lisenfeld2016}, thus the outlier $\delta$-values likely include a contribution from these gravitational interactions in addition to potential gas compression from RPS.
\par
The excess molecular gas mass on the leading half is roughly matched by an excess in SFR (Fig.~\ref{fig:lt_ratio}c), for the majority of galaxies in the sample. The distribution of $\delta_\mathrm{SFR}$ in Fig.~\ref{fig:lt_ratio}c is also skewed to values greater than one with a median that is greater than one at $\sim\!3\sigma$ significance. As is shown in Fig.~\ref{fig:lt_ratio}d, the increase in \htwo{} mass and SFR are such that, on average, the molecular gas depletion time on the leading half is not elevated or suppressed relative to the trailing half, and instead scatters around the one-to-one line (i.e.\ $\delta_{t_\mathrm{dep}} = 1$). We further highlight the connection between $\delta_\mathrm{H_2}$ and $\delta_\mathrm{SFR}$ in Fig.~\ref{fig:delta_h2_sfr} where we plot the two against one another. The correlation between $\delta_\mathrm{H_2}$ and $\delta_\mathrm{SFR}$ is apparent by-eye and Spearman's rank test reinforces this with a correlation coefficient of $r_s = 0.61$ ($p\text{-value} = 0.01$). The one outlier to the trend in Fig.~\ref{fig:delta_h2_sfr} is NGC4396, which shows a clear excess in SFR on the leading half but a small deficit of molecular gas (relative to the trailing half). This is primarily driven by a bright star-forming region on the very leading edge of NGC4396. The results from Figs~\ref{fig:lt_ratio}d and \ref{fig:delta_h2_sfr} are consistent with recent work from \citet{roberts2022_lofar_manga} who report evidence for enhanced CO emission and star formation on the leading half of the Coma cluster jellyfish galaxy, IC3949. Similar to this work, \citeauthor{roberts2022_lofar_manga} find that the excess CO emission and SFR are in proportion such that the depletion time in the region of enhanced star formation in IC3949 matches the typical depletion time over the rest of the disc.
\par
In Fig.~\ref{fig:lt_ratio}e we find that $\delta_\bigstar$ is centred on one with relatively little scatter. This result may be physically meaningful, in the sense that RPS is not expected to strongly perturb a galaxy's stellar disc. Therefore, if it is ram pressure driving the molecular gas and SFR anisotropies, the $\delta_\bigstar$ distribution in Fig.~\ref{fig:lt_ratio}e is consistent with this scenario. That said, we stress that the galaxy centres used to divide the leading and trailing halves are determined from the stellar mass centroid for each galaxy and therefore this approach will naturally tend towards $\delta_\bigstar$ values with small magnitudes.

\begin{figure}[!ht]
    \centering
    \includegraphics[width=\columnwidth]{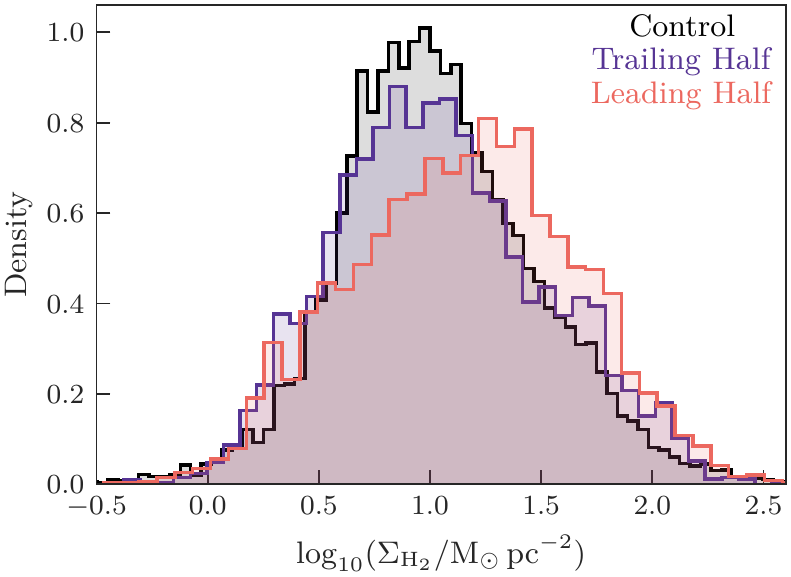}
    \caption{\htwo{} surface density distributions for detected pixels in control sample galaxies (black) as well as detected pixels on the leading half (orange) and trailing half (purple) of \hi{}-tailed galaxies. \htwo{} surface density distributions are taken from the $\mathrm{720}\text{-}\mathrm{pc}$ resolution maps. Bin widths are set according to "Knuth's rule" \citep{knuth2006} using the histogram functionality within \texttt{astropy.visualization}. Compared to the control and trailing half, pixels on the leading half are skewed towards higher \htwo{} surface densities.}
    \label{fig:gas_density_dist_lt}
\end{figure}
An important question raised by the results of this section is whether the excess molecular gas mass observed on the leading half is a result of gas compression leading to enhanced CO emission, or due to molecular gas being preferentially removed from the trailing half. For some of the galaxies in our sample: NGC4402, NGC4501. NGC4522, NGC4654, evidence for molecular gas compression has already been reported in literature \citep{chung2014,nehlig2016,lee2017,cramer2020,lizee2021}. In Fig~\ref{fig:gas_density_dist_lt} we show the distributions of \htwo{} surface density for the pixels in control galaxies as well as pixels on the leading half and trailing half of \hi{}-tailed galaxies. The distributions for the control and trailing half pixels are very similar, whereas the distribution for leading half pixels is skewed towards larger values (by $\sim\!0.5-1\,\mathrm{dex}$ in terms of the peaks of the distributions). We take this as evidence that the \htwo{} excess from Fig.~\ref{fig:lt_ratio} is tracing enhanced CO emission, potentially due to compression from ram pressure, as opposed to a deficit of molecular gas on the trailing half. Though we do note that there is almost certainly a combination of gas compression and direct molecular gas removal simultaneously affecting the \htwo{} distributions in cluster galaxies.  As shown in previous VERTICO papers, the direct stripping aspect most strongly impacts low-density molecular gas \citep{zabel2022,watts2023}. The enhanced \htwo{} surface densities in Figs~\ref{fig:lt_ratio} and \ref{fig:gas_density_dist_lt} are also consistent with results from \citet{brown2023}, who find that `early-RPS' galaxies in VERTICO have larger molecular gas densities than what is seen for a non-cluster control sample.
\par
The fact that the excess molecular gas mass on the leading side coincides with a similar excess in SFR, supports the framework where the enhanced SFRs observed for galaxies undergoing RPS \citep[e.g.][]{dressler1983,ebeling2014,vulcani2018_sf,roberts2020,vulcani2020,roberts2021_LOFARclust,lee2022_enhanced_sfr} are catalyzed by gas compression in the ISM \citep[e.g.][]{gavazzi2001,lee2016,boselli2021_ic3476,roberts2022_perseus,hess2022,roberts2022_lofar_manga}. Though we note that these star formation enhancements, if common, are likely short-lived \citep{boselli2022}. These effects do not drive a systematic difference in molecular gas depletion times between the leading and trailing halves, as can be seen in Fig.~\ref{fig:lt_ratio}d. This is more broadly consistent with previous works that have shown that the Kennicutt-Schmidt (KS, \citealt{schmidt1959,kennicutt1998}) relationship for cluster galaxies does not differ strongly from the KS relation for isolated galaxies \citep{nehlig2016,zabel2020,jimenez-donaire2022}. Thus perturbations within the environment of a galaxy cluster can alter the distribution of molecular gas within galaxy discs, but these perturbations do not appear to alter the star formation law itself, at least for cluster galaxies that are still relatively gas rich. Cluster galaxies that are very \hi{}-deficient (i.e.\ are at a very advanced stage of environmental evolution) do show unusually long depletion times \citep{jimenez-donaire2022,brown2023}.

\section{Conclusion} \label{sec:conclusion}

In this paper we analyze the spatial distributions of \htwo{} and \hi{} gas surface densities in Virgo cluster galaxies with observations from the VERTICO and VIVA surveys. Our approach is twofold: we first consider a quantitative measure of rotational asymmetry, $A_\mathrm{mod}$, and compare \hi{} and \htwo{} asymmetries across different stages of environmental influence, and second, we divide \hi{}-tailed galaxies into a leading and trailing half in order to test for the presence of ram-pressure induced gas compression and star formation. The main scientific results from this work are as follows:
\begin{itemize}
    \itemsep0.5em
    \item The quantitative \hi{} asymmetries measured for Virgo galaxies follow the \hi{} stripping classes from \citet{yoon2017}; namely, \hi{}-tailed galaxies have the largest asymmetries compared to \hi{}-normal and \hi{}-truncated galaxies. On the whole, Virgo galaxies have \hi{} asymmetries that are larger than control sample galaxies by $40 \pm 10$ per cent, on average (Fig.~\ref{fig:asym_HIclass}a).
    
    \item There is less difference in quantitative \htwo{} asymmetries for Virgo galaxies across \hi{} stripping classes. Virgo galaxies have \htwo{} asymmetries that are, on average, larger than the control sample by $20 \pm 10$ per cent, but there are not strong differences in the \htwo{} asymmetries for \hi{}-normal, \hi{}-tailed, and \hi{}-truncated galaxies (Fig.~\ref{fig:asym_HIclass}b).
    
    \item On a galaxy-by-galaxy basis there is a weak correlation between \hi{} and \htwo{} asymmetries at matched resolution. Control sample galaxies show slightly larger \htwo{} asymmetries than \hi{} asymmetries, whereas Virgo cluster galaxies have larger \hi{} asymmetries than \htwo{} asymmetries. For galaxies with the most asymmetric gas distributions, we observe a stronger correlation between \hi{} and \htwo{} asymmetries (Fig.~\ref{fig:asym_hi_co}).
    
    \item We find evidence for enhanced \htwo{} content on the leading halves of \hi{}-tailed galaxies compared to both the trailing halves of those same galaxies and galaxies in the control sample (Figs \ref{fig:lt_ratio}b and \ref{fig:gas_density_dist_lt}).
    
    \item The leading halves of \hi{}-tailed galaxies have larger star formation rates than the trailing halves. The enhanced star formation is in proportion with the molecular gas such that the molecular gas depletion time, on average, does not differ between the leading and trailing halves (Figs \ref{fig:lt_ratio}d and \ref{fig:delta_h2_sfr}).
\end{itemize}

As demonstrated and discussed throughout this paper, there is an impact from the cluster environment on the spatial distribution of cold gas in Virgo satellite galaxies. The results from this work are consistent with the following general, schematic framework for the evolution of cold gas in cluster galaxies:
\par
The asymmetry of atomic gas appears to increase at an early stage of environmental evolution, such that even cluster galaxies with little-to-no \hi{} deficiency still show slightly larger \hi{} asymmetries than non-cluster galaxies. As environmental perturbations (e.g. ram pressure) become stronger, some (maybe most or even all) star-forming satellites develop an extended \hi{} tail as gas is removed from the galaxy.  This leads to a large quantitative \hi{} asymmetry. As atomic gas continues to be depleted, the measured \hi{} asymmetry decreases as the \hi{} disc becomes truncated and centrally concentrated relative to the stellar disc.  At this post-stripping phase, the quantitative \hi{} asymmetry can be similar to values for non-\hi{}-deficient, or even non-cluster, galaxies. Thus measured atomic gas asymmetry itself is not a monotonic tracer of environmental influence, and for a full picture it should be coupled with measurements of \hi{} size or deficiency.
\par
For molecular gas, the results of this work are consistent with a small increase in \htwo{} asymmetry for cluster galaxies. For the most advanced stages of environmental evolution (i.e.\ \hi{}-tailed and \hi{}-truncated in this work), there is a correlation between \hi{} and \htwo{} asymmetry, with the \hi{} asymmetry almost always being the larger of the two. Thus there is an evironmental effect on \htwo{} morphologies, but it is of a smaller magnitude than that for \hi{}. For some \hi{}-tailed galaxies, the perturbations to \htwo{} manifest as excess CO emission on the leading half of the galaxy, potentially due to gas compression. This can trigger star formation, which likely contributes to the large SFRs that have been observed for galaxies undergoing RPS \citep[e.g.][]{dressler1983,vulcani2018_sf,vulcani2020,roberts2021_LOFARclust,lee2022_enhanced_sfr}.
\par
The above framework is consistent with the results of this work, though our sample size is still relatively modest and this work only probes a single galaxy cluster. As the number of resolved cold-gas observations of cluster galaxies continues to increase, the applicability of this picture in general will become better constrained.
\par
As ever, cold-gas measurements at higher spatial resolutions will further advance the results presented in this work. This will give more precise measurements of quantitative asymmetries (i.e.\ measuring $A_\mathrm{mod}$ closer to the intrinsic value for each galaxy), but also in the case of cloud-scale CO measurements, could constrain the impact of ram pressure and other environmental mechanisms on individual star-forming regions.  In principle this is already possible for one third of the \hi{}-tailed galaxies from this work (NGC4254, NGC4298, NGC4424, NGC4535, NGC4654, NGC4694) that have cloud-scale CO data available from PHANGS \citep{leroy2021_phangs}, though we leave this analysis for a future work.
\par
Given the elevated molecular gas masses observed on the leading half of \hi{}-tailed galaxies in this work, it should be explored whether or not this asymmetry is also reflected by the distribution of dense molecular gas tracers (i.e.\ HCN, HCO$^+$). This would be expected in the framework of ram pressure gas compression, and has been already observed for IC3949 in the Coma cluster \citep{roberts2022_lofar_manga}.  For one galaxy from the sample, NGC4654, HCN (1-0) observations have been recently obtained from the IRAM 30m telescope (P.I. M. Jim{\'e}nez-Donaire), which will allow this to be tested directly.

\begin{acknowledgements}
The authors thank the anonymous referee for providing a timely and constructive report, which has resulted in an improved manuscript.  We also thank Seona Lee for sharing the \hi{} tail orientations for the Virgo galaxies in this work. This work was carried out as part of the VERTICO collaboration. IDR acknowledges support from the Natural Sciences and Engineering Research Council of Canada and the ERC Starting Grant Cluster Web 804208 (PI, R.J.\ van Weeren). TB acknowledges support from the National Research Council of Canada via the Plaskett Fellowship of the Dominion Astrophysical Observatory. The financial assistance of the National Research Foundation (NRF) towards this research is hereby acknowledged by NZ. Opinions expressed and conclusions arrived at, are those of the author and are not necessarily to be attributed to the NRF. NZ is supported through the South African Research Chairs Initiative of
the Department of Science and Technology and National Research Foundation. CDW acknowledges support from the Natural Sciences and Engineering Research Council of Canada and the Canada Research Chairs program. LCP and KS acknowledge support from the Natural Sciences and Engineering Research Council of Canada. ABW and LC acknowledge support from the Australian Research Council Discovery Project funding scheme (DP210100337). TAD acknowledges support from the UK Science and Technology Facilities Council through grants ST/S00033X/1 and ST/W000830/1. BL acknowledges the support from the Korea Astronomy and Space Science Institute grant funded by the Korea government (MSIT) (Project No. 2022-1-840-05).  ARHS acknowledges receipt of the Jim Buckee Fellowship at ICRAR-UWA.  VV acknowledges support from the scholarship ANID-FULBRIGHT BIO 2016 - 56160020, funding from NRAO Student Observing Support (SOS)
- SOSPA7-014, and partial support from NSF-AST2108140. Parts of this research were supported by the Australian Research Council Centre of Excellence for All Sky Astrophysics in 3 Dimensions (ASTRO 3D), through project number CE170100013. We acknowledge the usage of the HyperLeda database (\url{http://leda.univ-lyon1.fr}).
\\[0.5em]
This paper makes use of the following ALMA data: \\[0.25em]
ADS/JAO.ALMA \href{https://almascience.nrao.edu/aq/?result_view=projects&projectCode=\%222019.1.00763.L\%22}{\#2019.1.00763.L} \\
ADS/JAO.ALMA \href{https://almascience.nrao.edu/aq/?result_view=projects&projectCode=\%222017.1.00886.L\%22}{\#2017.1.00886.L}. \\
ADS/JAO.ALMA \href{https://almascience.nrao.edu/aq/?result_view=projects&projectCode=\%222016.1.00912.S\%22}{\#2016.1.00912.S} \\
ADS/JAO.ALMA \href{https://almascience.nrao.edu/aq/?result_view=projects&projectCode=\%222015.1.00956.S\%22}{\#2015.1.00956.S}. \\[0.25em]
ALMA is a partnership of ESO (representing its member states), NSF (USA) and NINS (Japan), together with NRC (Canada), MOST and ASIAA (Taiwan), and KASI (Republic of Korea), in cooperation with the Republic of Chile. The Joint ALMA Observatory is operated by ESO, AUI/NRAO and NAOJ. In addition, publications from NA authors must include the standard NRAO acknowledgement: The National Radio Astronomy Observatory is a facility of the National Science Foundation operated under cooperative agreement by Associated Universities, Inc.
\end{acknowledgements}



%
%

\bibliographystyle{aa}
\bibliography{main}

\begin{thebibliography}{152}
\expandafter\ifx\csname natexlab\endcsname\relax\def\natexlab#1{#1}\fi

\bibitem[{{Abraham} {et~al.}(1996){Abraham}, {Tanvir}, {Santiago}, {Ellis},
  {Glazebrook}, \& {van den Bergh}}]{abraham1996}
{Abraham}, R.~G., {Tanvir}, N.~R., {Santiago}, B.~X., {et~al.} 1996, MNRAS,
  279, L47

\bibitem[{{Anderson} \& {Darling}(1952)}]{anderson1952}
{Anderson}, T.~W. \& {Darling}, D.~A. 1952, The Annals of Mathematical
  Statistics, 23, 193

\bibitem[{{Bekki}(2014)}]{bekki2014}
{Bekki}, K. 2014, MNRAS, 438, 444

\bibitem[{{Bershady} {et~al.}(2000){Bershady}, {Jangren}, \&
  {Conselice}}]{bershady2000}
{Bershady}, M.~A., {Jangren}, A., \& {Conselice}, C.~J. 2000, AJ, 119, 2645

\bibitem[{{Bigiel} {et~al.}(2011){Bigiel}, {Leroy}, {Walter}, {Brinks}, {de
  Blok}, {Kramer}, {Rix}, {Schruba}, {Schuster}, {Usero}, \&
  {Wiesemeyer}}]{bigiel2011}
{Bigiel}, F., {Leroy}, A.~K., {Walter}, F., {et~al.} 2011, ApJL, 730, L13

\bibitem[{{Bilimogga} {et~al.}(2022){Bilimogga}, {Oman}, {Verheijen}, \& {van
  der Hulst}}]{bilimogga2022}
{Bilimogga}, P.~V., {Oman}, K.~A., {Verheijen}, M. A.~W., \& {van der Hulst},
  T. 2022, MNRAS, 513, 5310

\bibitem[{{Blanton} {et~al.}(2017){Blanton}, {Bershady}, {Abolfathi},
  {Albareti}, {Allende Prieto}, {Almeida}, {Alonso-Garc{\'\i}a}, {Anders},
  {Anderson}, {Andrews}, {Aquino-Ort{\'\i}z}, {Arag{\'o}n-Salamanca},
  {Argudo-Fern{\'a}ndez}, {Armengaud}, {Aubourg}, {Avila-Reese}, {Badenes},
  {Bailey}, {Barger}, {Barrera-Ballesteros}, {Bartosz}, {Bates}, {Baumgarten},
  {Bautista}, {Beaton}, {Beers}, {Belfiore}, {Bender}, {Berlind}, {Bernardi},
  {Beutler}, {Bird}, {Bizyaev}, {Blanc}, {Blomqvist}, {Bolton}, {Boquien},
  {Borissova}, {van den Bosch}, {Bovy}, {Brandt}, {Brinkmann}, {Brownstein},
  {Bundy}, {Burgasser}, {Burtin}, {Busca}, {Cappellari}, {Delgado Carigi},
  {Carlberg}, {Carnero Rosell}, {Carrera}, {Chanover}, {Cherinka}, {Cheung},
  {G{\'o}mez Maqueo Chew}, {Chiappini}, {Choi}, {Chojnowski}, {Chuang},
  {Chung}, {Cirolini}, {Clerc}, {Cohen}, {Comparat}, {da Costa}, {Cousinou},
  {Covey}, {Crane}, {Croft}, {Cruz-Gonzalez}, {Garrido Cuadra}, {Cunha},
  {Damke}, {Darling}, {Davies}, {Dawson}, {de la Macorra}, {Dell'Agli}, {De
  Lee}, {Delubac}, {Di Mille}, {Diamond-Stanic}, {Cano-D{\'\i}az}, {Donor},
  {Downes}, {Drory}, {du Mas des Bourboux}, {Duckworth}, {Dwelly}, {Dyer},
  {Ebelke}, {Eigenbrot}, {Eisenstein}, {Emsellem}, {Eracleous}, {Escoffier},
  {Evans}, {Fan}, {Fern{\'a}ndez-Alvar}, {Fernandez-Trincado}, {Feuillet},
  {Finoguenov}, {Fleming}, {Font-Ribera}, {Fredrickson}, {Freischlad},
  {Frinchaboy}, {Fuentes}, {Galbany}, {Garcia-Dias},
  {Garc{\'\i}a-Hern{\'a}ndez}, {Gaulme}, {Geisler}, {Gelfand},
  {Gil-Mar{\'\i}n}, {Gillespie}, {Goddard}, {Gonzalez-Perez}, {Grabowski},
  {Green}, {Grier}, {Gunn}, {Guo}, {Guy}, {Hagen}, {Hahn}, {Hall}, {Harding},
  {Hasselquist}, {Hawley}, {Hearty}, {Gonzalez Hern{\'a}ndez}, {Ho}, {Hogg},
  {Holley-Bockelmann}, {Holtzman}, {Holzer}, {Huehnerhoff}, {Hutchinson},
  {Hwang}, {Ibarra-Medel}, {da Silva Ilha}, {Ivans}, {Ivory}, {Jackson},
  {Jensen}, {Johnson}, {Jones}, {J{\"o}nsson}, {Jullo}, {Kamble}, {Kinemuchi},
  {Kirkby}, {Kitaura}, {Klaene}, {Knapp}, {Kneib}, {Kollmeier}, {Lacerna},
  {Lane}, {Lang}, {Law}, {Lazarz}, {Lee}, {Le Goff}, {Liang}, {Li}, {Li},
  {Lian}, {Lima}, {Lin}, {Lin}, {Bertran de Lis}, {Liu}, {de Icaza Lizaola},
  {Long}, {Lucatello}, {Lundgren}, {MacDonald}, {Deconto Machado}, {MacLeod},
  {Mahadevan}, {Geimba Maia}, {Maiolino}, {Majewski}, {Malanushenko},
  {Malanushenko}, {Manchado}, {Mao}, {Maraston}, {Marques-Chaves}, {Masseron},
  {Masters}, {McBride}, {McDermid}, {McGrath}, {McGreer}, {Medina Pe{\~n}a},
  {Melendez}, {Merloni}, {Merrifield}, {Meszaros}, {Meza}, {Minchev},
  {Minniti}, {Miyaji}, {More}, {Mulchaey}, {M{\"u}ller-S{\'a}nchez}, {Muna},
  {Munoz}, {Myers}, {Nair}, {Nandra}, {Correa do Nascimento}, {Negrete},
  {Ness}, {Newman}, {Nichol}, {Nidever}, {Nitschelm}, {Ntelis}, {O'Connell},
  {Oelkers}, {Oravetz}, {Oravetz}, {Pace}, {Padilla}, {Palanque-Delabrouille},
  {Alonso Palicio}, {Pan}, {Parejko}, {Parikh}, {P{\^a}ris}, {Park}, {Patten},
  {Peirani}, {Pellejero-Ibanez}, {Penny}, {Percival}, {Perez-Fournon},
  {Petitjean}, {Pieri}, {Pinsonneault}, {Pisani}, {Poleski}, {Prada},
  {Prakash}, {Queiroz}, {Raddick}, {Raichoor}, {Barboza Rembold}, {Richstein},
  {Riffel}, {Riffel}, {Rix}, {Robin}, {Rockosi}, {Rodr{\'\i}guez-Torres},
  {Roman-Lopes}, {Rom{\'a}n-Z{\'u}{\~n}iga}, {Rosado}, {Ross}, {Rossi}, {Ruan},
  {Ruggeri}, {Rykoff}, {Salazar-Albornoz}, {Salvato}, {S{\'a}nchez}, {Aguado},
  {S{\'a}nchez-Gallego}, {Santana}, {Santiago}, {Sayres}, {Schiavon}, {da Silva
  Schimoia}, {Schlafly}, {Schlegel}, {Schneider}, {Schultheis}, {Schuster},
  {Schwope}, {Seo}, {Shao}, {Shen}, {Shetrone}, {Shull}, {Simon}, {Skinner},
  {Skrutskie}, {Slosar}, {Smith}, {Sobeck}, {Sobreira}, {Somers}, {Souto},
  {Stark}, {Stassun}, {Stauffer}, {Steinmetz}, {Storchi-Bergmann},
  {Streblyanska}, {Stringfellow}, {Su{\'a}rez}, {Sun}, {Suzuki}, {Szigeti},
  {Taghizadeh-Popp}, {Tang}, {Tao}, {Tayar}, {Tembe}, {Teske}, {Thakar},
  {Thomas}, {Thompson}, {Tinker}, {Tissera}, {Tojeiro}, {Hernandez Toledo}, {de
  la Torre}, {Tremonti}, {Troup}, {Valenzuela}, {Martinez Valpuesta},
  {Vargas-Gonz{\'a}lez}, {Vargas-Maga{\~n}a}, {Vazquez}, {Villanova}, {Vivek},
  {Vogt}, {Wake}, {Walterbos}, {Wang}, {Weaver}, {Weijmans}, {Weinberg},
  {Westfall}, {Whelan}, {Wild}, {Wilson}, {Wood-Vasey}, {Wylezalek}, {Xiao},
  {Yan}, {Yang}, {Ybarra}, {Y{\`e}che}, {Zakamska}, {Zamora}, {Zarrouk},
  {Zasowski}, {Zhang}, {Zhao}, {Zheng}, {Zheng}, {Zhou}, {Zhou}, {Zhu},
  {Zoccali}, \& {Zou}}]{blanton2017}
{Blanton}, M.~R., {Bershady}, M.~A., {Abolfathi}, B., {et~al.} 2017, AJ, 154,
  28

\bibitem[{{Bolatto} {et~al.}(2008){Bolatto}, {Leroy}, {Rosolowsky}, {Walter},
  \& {Blitz}}]{bolatto2008}
{Bolatto}, A.~D., {Leroy}, A.~K., {Rosolowsky}, E., {Walter}, F., \& {Blitz},
  L. 2008, ApJ, 686, 948

\bibitem[{{Boselli} {et~al.}(2016{\natexlab{a}}){Boselli}, {Boissier}, {Voyer},
  {Ferrarese}, {Consolandi}, {Cortese}, {C{\^o}t{\'e}}, {Cuillandre},
  {Gavazzi}, {Gwyn}, {Heinis}, {Ilbert}, {MacArthur}, \&
  {Roehlly}}]{boselli2016_galex}
{Boselli}, A., {Boissier}, S., {Voyer}, E., {et~al.} 2016{\natexlab{a}}, A\&A,
  585, A2

\bibitem[{{Boselli} {et~al.}(2014{\natexlab{a}}){Boselli}, {Cortese}, \&
  {Boquien}}]{boselli2014_co}
{Boselli}, A., {Cortese}, L., \& {Boquien}, M. 2014{\natexlab{a}}, A\&A, 564,
  A65

\bibitem[{{Boselli} {et~al.}(2014{\natexlab{b}}){Boselli}, {Cortese},
  {Boquien}, {Boissier}, {Catinella}, {Gavazzi}, {Lagos}, \&
  {Saintonge}}]{boselli2014}
{Boselli}, A., {Cortese}, L., {Boquien}, M., {et~al.} 2014{\natexlab{b}}, A\&A,
  564, A67

\bibitem[{{Boselli} {et~al.}(2016{\natexlab{b}}){Boselli}, {Cuillandre},
  {Fossati}, {Boissier}, {Bomans}, {Consolandi}, {Anselmi}, {Cortese},
  {C{\^o}t{\'e}}, {Durrell}, {Ferrarese}, {Fumagalli}, {Gavazzi}, {Gwyn},
  {Hensler}, {Sun}, \& {Toloba}}]{boselli2016_ngc4569}
{Boselli}, A., {Cuillandre}, J.~C., {Fossati}, M., {et~al.} 2016{\natexlab{b}},
  \aap, 587, A68

\bibitem[{{Boselli} {et~al.}(2018){Boselli}, {Fossati}, {Ferrarese},
  {Boissier}, {Consolandi}, {Longobardi}, {Amram}, {Balogh}, {Barmby},
  {Boquien}, {Boulanger}, {Braine}, {Buat}, {Burgarella}, {Combes}, {Contini},
  {Cortese}, {C{\^o}t{\'e}}, {C{\^o}t{\'e}}, {Cuilland re}, {Drissen},
  {Epinat}, {Fumagalli}, {Gallagher}, {Gavazzi}, {Gomez-Lopez}, {Gwyn},
  {Harris}, {Hensler}, {Koribalski}, {Marcelin}, {McConnachie},
  {Miville-Deschenes}, {Navarro}, {Patton}, {Peng}, {Plana}, {Prantzos},
  {Robert}, {Roediger}, {Roehlly}, {Russeil}, {Salome}, {Sanchez-Janssen},
  {Serra}, {Spekkens}, {Sun}, {Taylor}, {Tonnesen}, {Vollmer}, {Willis},
  {Wozniak}, {Burdullis}, {Devost}, {Mahoney}, {Manset}, {Petric}, {Prunet}, \&
  {Withington}}]{boselli2018}
{Boselli}, A., {Fossati}, M., {Ferrarese}, L., {et~al.} 2018, A\&A, 614, A56

\bibitem[{{Boselli} {et~al.}(2015){Boselli}, {Fossati}, {Gavazzi}, {Ciesla},
  {Buat}, {Boissier}, \& {Hughes}}]{boselli2015}
{Boselli}, A., {Fossati}, M., {Gavazzi}, G., {et~al.} 2015, A\&A, 579, A102

\bibitem[{{Boselli} {et~al.}(2023){Boselli}, {Fossati}, {Roediger}, {Boquien},
  {Fumagalli}, {Balogh}, {Boissier}, {Braine}, {Ciesla}, {C{\^o}t{\'e}},
  {Cuillandre}, {Ferrarese}, {Gavazzi}, {Gwyn}, {Junais}, {Hensler},
  {Longobardi}, \& {Sun}}]{boselli2022}
{Boselli}, A., {Fossati}, M., {Roediger}, J., {et~al.} 2023, A\&A, 669, A73

\bibitem[{{Boselli} {et~al.}(2022){Boselli}, {Fossati}, \&
  {Sun}}]{boselli2022_review}
{Boselli}, A., {Fossati}, M., \& {Sun}, M. 2022, A\&AR, 30, 3

\bibitem[{{Boselli} {et~al.}(1997){Boselli}, {Gavazzi}, {Lequeux}, {Buat},
  {Casoli}, {Dickey}, \& {Donas}}]{boselli1997}
{Boselli}, A., {Gavazzi}, G., {Lequeux}, J., {et~al.} 1997, A\&A, 327, 522

\bibitem[{{Boselli} {et~al.}(2021){Boselli}, {Lupi}, {Epinat}, {Amram},
  {Fossati}, {Anderson}, {Boissier}, {Boquien}, {Consolandi}, {C{\^o}t{\'e}},
  {Cuillandre}, {Ferrarese}, {Galbany}, {Gavazzi}, {G{\'o}mez-L{\'o}pez},
  {Gwyn}, {Hensler}, {Hutchings}, {Kuncarayakti}, {Longobardi}, {Peng},
  {Plana}, {Postma}, {Roediger}, {Roehlly}, {Schimd}, {Trinchieri}, \&
  {Vollmer}}]{boselli2021_ic3476}
{Boselli}, A., {Lupi}, A., {Epinat}, B., {et~al.} 2021, A\&A, 646, A139

\bibitem[{{Bravo-Alfaro} {et~al.}(2000){Bravo-Alfaro}, {Cayatte}, {van Gorkom},
  \& {Balkowski}}]{bravo-alfaro2000}
{Bravo-Alfaro}, H., {Cayatte}, V., {van Gorkom}, J.~H., \& {Balkowski}, C.
  2000, AJ, 119, 580

\bibitem[{{Briggs}(1995)}]{briggs1995}
{Briggs}, D.~S. 1995, in American Astronomical Society Meeting Abstracts, Vol.
  187, American Astronomical Society Meeting Abstracts, 112.02

\bibitem[{{Brown} {et~al.}(2017){Brown}, {Catinella}, {Cortese}, {Lagos},
  {Dav{\'e}}, {Kilborn}, {Haynes}, {Giovanelli}, \&
  {Rafieferantsoa}}]{brown2017}
{Brown}, T., {Catinella}, B., {Cortese}, L., {et~al.} 2017, MNRAS, 466, 1275

\bibitem[{{Brown} {et~al.}(submitted){Brown}, {Roberts}, {Thorp}, {Ellison},
  {Zabel}, {Wilson}, {Bah{\'e}}, {Bisaria}, {Bolatto}, {Boselli}, {Chung},
  {Cortese}, {Catinella}, {Davis}, {Jim{\'e}nez-Donaire}, {Lagos}, {Lee},
  {Parker}, {Smith}, {Spekkens}, {Stevens}, {Villanueva}, \& B.}]{brown2023}
{Brown}, T., {Roberts}, I.~D., {Thorp}, M., {et~al.} submitted

\bibitem[{{Brown} {et~al.}(2021){Brown}, {Wilson}, {Zabel}, {Davis}, {Boselli},
  {Chung}, {Ellison}, {Lagos}, {Stevens}, {Cortese}, {Bah{\'e}}, {Bisaria},
  {Bolatto}, {Cashmore}, {Catinella}, {Chown}, {Diemer}, {Elahi}, {Hani},
  {Jim{\'e}nez-Donaire}, {Lee}, {Leidig}, {Mok}, {Olsen}, {Parker}, {Roberts},
  {Smith}, {Spekkens}, {Thorp}, {Tonnesen}, {Vienneau}, {Villanueva}, {Vogel},
  {Wadsley}, {Welker}, \& {Yoon}}]{brown2021}
{Brown}, T., {Wilson}, C.~D., {Zabel}, N., {et~al.} 2021, ApJS, 257, 21

\bibitem[{{Catinella} {et~al.}(2018){Catinella}, {Saintonge}, {Janowiecki},
  {Cortese}, {Dav{\'e}}, {Lemonias}, {Cooper}, {Schiminovich}, {Hummels},
  {Fabello}, {Ger{\'e}b}, {Kilborn}, \& {Wang}}]{catinella2018}
{Catinella}, B., {Saintonge}, A., {Janowiecki}, S., {et~al.} 2018, MNRAS, 476,
  875

\bibitem[{{Catinella} {et~al.}(2010){Catinella}, {Schiminovich}, {Kauffmann},
  {Fabello}, {Wang}, {Hummels}, {Lemonias}, {Moran}, {Wu}, {Giovanelli},
  {Haynes}, {Heckman}, {Basu-Zych}, {Blanton}, {Brinchmann}, {Budav{\'a}ri},
  {Gon{\c{c}}alves}, {Johnson}, {Kennicutt}, {Madore}, {Martin}, {Rich},
  {Tacconi}, {Thilker}, {Wild}, \& {Wyder}}]{catinella2010}
{Catinella}, B., {Schiminovich}, D., {Kauffmann}, G., {et~al.} 2010, MNRAS,
  403, 683

\bibitem[{{Chabrier}(2003)}]{chabrier2003}
{Chabrier}, G. 2003, PASP, 115, 763

\bibitem[{{Chung} {et~al.}(2009){Chung}, {van Gorkom}, {Kenney}, {Crowl}, \&
  {Vollmer}}]{chung2009}
{Chung}, A., {van Gorkom}, J.~H., {Kenney}, J.~D.~P., {Crowl}, H., \&
  {Vollmer}, B. 2009, AJ, 138, 1741

\bibitem[{{Chung} {et~al.}(2007){Chung}, {van Gorkom}, {Kenney}, \&
  {Vollmer}}]{chung2007}
{Chung}, A., {van Gorkom}, J.~H., {Kenney}, J.~D.~P., \& {Vollmer}, B. 2007,
  ApJl, 659, L115

\bibitem[{{Chung} \& {Kim}(2014)}]{chung2014}
{Chung}, E.~J. \& {Kim}, S. 2014, PASJ, 66, 11

\bibitem[{{Conselice}(2003)}]{conselice2003_cas}
{Conselice}, C.~J. 2003, ApJS, 147, 1

\bibitem[{{Conselice} {et~al.}(2000){Conselice}, {Bershady}, \&
  {Jangren}}]{conselice2000}
{Conselice}, C.~J., {Bershady}, M.~A., \& {Jangren}, A. 2000, ApJ, 529, 886

\bibitem[{{Cort{\'e}s} {et~al.}(2006){Cort{\'e}s}, {Kenney}, \&
  {Hardy}}]{cortes2006}
{Cort{\'e}s}, J.~R., {Kenney}, J. D.~P., \& {Hardy}, E. 2006, AJ, 131, 747

\bibitem[{{Cortese} {et~al.}(2012){Cortese}, {Boissier}, {Boselli}, {Bendo},
  {Buat}, {Davies}, {Eales}, {Heinis}, {Isaak}, \& {Madden}}]{cortese2012}
{Cortese}, L., {Boissier}, S., {Boselli}, A., {et~al.} 2012, A\&A, 544, A101

\bibitem[{{Cortese} {et~al.}(2021){Cortese}, {Catinella}, \&
  {Smith}}]{cortese2021}
{Cortese}, L., {Catinella}, B., \& {Smith}, R. 2021, PASA, 38, e035

\bibitem[{{Cortese} {et~al.}(2007){Cortese}, {Marcillac}, {Richard},
  {Bravo-Alfaro}, {Kneib}, {Rieke}, {Covone}, {Egami}, {Rigby}, {Czoske}, \&
  {Davies}}]{cortese2007}
{Cortese}, L., {Marcillac}, D., {Richard}, J., {et~al.} 2007, MNRAS, 376, 157

\bibitem[{{Cramer} {et~al.}(2020){Cramer}, {Kenney}, {Cortes}, {Cortes P.~C.},
  {Vlahakis}, {J{\'a}chym}, {Pompei}, \& {Rubio}}]{cramer2020}
{Cramer}, W.~J., {Kenney}, J.~D.~P., {Cortes}, J.~R., {et~al.} 2020, ApJ, 901,
  95

\bibitem[{{Cramer} {et~al.}(2021){Cramer}, {Kenney}, {Tonnesen}, {Smith},
  {Wong}, {J{\'a}chym}, {Cort{\'e}s}, {Cort{\'e}s}, \& {Wu}}]{cramer2021}
{Cramer}, W.~J., {Kenney}, J.~D.~P., {Tonnesen}, S., {et~al.} 2021, ApJ, 921,
  22

\bibitem[{{Croton} {et~al.}(2005){Croton}, {Farrar}, {Norberg}, {Colless},
  {Peacock}, {Baldry}, {Baugh}, {Bland-Hawthorn}, {Bridges}, {Cannon}, {Cole},
  {Collins}, {Couch}, {Dalton}, {De Propris}, {Driver}, {Efstathiou}, {Ellis},
  {Frenk}, {Glazebrook}, {Jackson}, {Lahav}, {Lewis}, {Lumsden}, {Maddox},
  {Madgwick}, {Peterson}, {Sutherland}, \& {Taylor}}]{croton2005}
{Croton}, D.~J., {Farrar}, G.~R., {Norberg}, P., {et~al.} 2005, MNRAS, 356,
  1155

\bibitem[{{Davies} {et~al.}(2019){Davies}, {Robotham}, {Lagos}, {Driver},
  {Stevens}, {Bah{\'e}}, {Alpaslan}, {Bremer}, {Brown}, {Brough},
  {Bland-Hawthorn}, {Cortese}, {Elahi}, {Grootes}, {Holwerda}, {Ludlow},
  {McGee}, {Owers}, \& {Phillipps}}]{davies2019}
{Davies}, L.~J.~M., {Robotham}, A.~S.~G., {Lagos}, C. d.~P., {et~al.} 2019,
  MNRAS, 483, 5444

\bibitem[{{Davis} {et~al.}(2022){Davis}, {Gensior}, {Bureau}, {Cappellari},
  {Choi}, {Elford}, {Kruijssen}, {Lelli}, {Liang}, {Liu}, {Ruffa}, {Saito},
  {Sarzi}, {Schruba}, \& {Williams}}]{davis2022}
{Davis}, T.~A., {Gensior}, J., {Bureau}, M., {et~al.} 2022, MNRAS, 512, 1522

\bibitem[{Deb {et~al.}(2023)Deb, Verheijen, \& van~der Hulst}]{deb2023}
Deb, T., Verheijen, M. A.~W., \& van~der Hulst, J.~M. 2023
  [\eprint[arXiv]{2303.10141}]

\bibitem[{{Dressler}(1980)}]{dressler1980}
{Dressler}, A. 1980, ApJ, 236, 351

\bibitem[{{Dressler} \& {Gunn}(1983)}]{dressler1983}
{Dressler}, A. \& {Gunn}, J.~E. 1983, ApJ, 270, 7

\bibitem[{{Ebeling} {et~al.}(2014){Ebeling}, {Stephenson}, \&
  {Edge}}]{ebeling2014}
{Ebeling}, H., {Stephenson}, L.~N., \& {Edge}, A.~C. 2014, ApJL, 781, L40

\bibitem[{{Engargiola} {et~al.}(2003){Engargiola}, {Plambeck}, {Rosolowsky}, \&
  {Blitz}}]{engargiola2003}
{Engargiola}, G., {Plambeck}, R.~L., {Rosolowsky}, E., \& {Blitz}, L. 2003,
  ApJS, 149, 343

\bibitem[{{Fossati} {et~al.}(2018){Fossati}, {Mendel}, {Boselli}, {Cuilland
  re}, {Vollmer}, {Boissier}, {Consolandi}, {Ferrarese}, {Gwyn}, {Amram},
  {Boquien}, {Buat}, {Burgarella}, {Cortese}, {C{\^o}t{\'e}}, {C{\^o}t{\'e}},
  {Durrell}, {Fumagalli}, {Gavazzi}, {Gomez-Lopez}, {Hensler}, {Koribalski},
  {Longobardi}, {Peng}, {Roediger}, {Sun}, \& {Toloba}}]{fossati2018}
{Fossati}, M., {Mendel}, J.~T., {Boselli}, A., {et~al.} 2018, A\&A, 614, A57

\bibitem[{{Fumagalli} {et~al.}(2009){Fumagalli}, {Krumholz}, {Prochaska},
  {Gavazzi}, \& {Boselli}}]{fumagalli2009}
{Fumagalli}, M., {Krumholz}, M.~R., {Prochaska}, J.~X., {Gavazzi}, G., \&
  {Boselli}, A. 2009, ApJ, 697, 1811

\bibitem[{{Gavazzi} {et~al.}(2001){Gavazzi}, {Boselli}, {Mayer},
  {Iglesias-Paramo}, {V{\'\i}lchez}, \& {Carrasco}}]{gavazzi2001}
{Gavazzi}, G., {Boselli}, A., {Mayer}, L., {et~al.} 2001, ApJL, 563, L23

\bibitem[{{Gavazzi} \& {Jaffe}(1987)}]{gavazzi1987}
{Gavazzi}, G. \& {Jaffe}, W. 1987, A\&A, 186, L1

\bibitem[{{Giese} {et~al.}(2016){Giese}, {van der Hulst}, {Serra}, \&
  {Oosterloo}}]{giese2016}
{Giese}, N., {van der Hulst}, T., {Serra}, P., \& {Oosterloo}, T. 2016, MNRAS,
  461, 1656

\bibitem[{{Gunn} \& {Gott}(1972)}]{gunn1972}
{Gunn}, J.~E. \& {Gott}, III, J.~R. 1972, ApJ, 176, 1

\bibitem[{{Haynes} {et~al.}(1998){Haynes}, {Hogg}, {Maddalena}, {Roberts}, \&
  {van Zee}}]{haynes1998}
{Haynes}, M.~P., {Hogg}, D.~E., {Maddalena}, R.~J., {Roberts}, M.~S., \& {van
  Zee}, L. 1998, AJ, 115, 62

\bibitem[{{Heald} {et~al.}(2011){Heald}, {J{\'o}zsa}, {Serra}, {Zschaechner},
  {Rand}, {Fraternali}, {Oosterloo}, {Walterbos}, {J{\"u}tte}, \&
  {Gentile}}]{heald2011}
{Heald}, G., {J{\'o}zsa}, G., {Serra}, P., {et~al.} 2011, A\&A, 526, A118

\bibitem[{{Hess} {et~al.}(2022){Hess}, {Kotulla}, {Chen}, {Carignan},
  {Gallagher}, {Jarrett}, \& {Kraan-Korteweg}}]{hess2022}
{Hess}, K.~M., {Kotulla}, R., {Chen}, H., {et~al.} 2022, A\&A, 668, A184

\bibitem[{{Holwerda} {et~al.}(2011){Holwerda}, {Pirzkal}, {de Blok},
  {Bouchard}, {Blyth}, {van der Heyden}, \& {Elson}}]{holwerda2011_things}
{Holwerda}, B.~W., {Pirzkal}, N., {de Blok}, W.~J.~G., {et~al.} 2011, MNRAS,
  416, 2401

\bibitem[{{J{\'a}chym} {et~al.}(2014){J{\'a}chym}, {Combes}, {Cortese}, {Sun},
  \& {Kenney}}]{jachym2014}
{J{\'a}chym}, P., {Combes}, F., {Cortese}, L., {Sun}, M., \& {Kenney}, J. D.~P.
  2014, ApJ, 792, 11

\bibitem[{{J{\'a}chym} {et~al.}(2019){J{\'a}chym}, {Kenney}, {Sun}, {Combes},
  {Cortese}, {Scott}, {Sivanandam}, {Brinks}, {Roediger}, {Palou{\v{s}}}, \&
  {Fumagalli}}]{jachym2019}
{J{\'a}chym}, P., {Kenney}, J. D.~P., {Sun}, M., {et~al.} 2019, ApJ, 883, 145

\bibitem[{{J{\'a}chym} {et~al.}(2017){J{\'a}chym}, {Sun}, {Kenney}, {Cortese},
  {Combes}, {Yagi}, {Yoshida}, {Palou{\v{s}}}, \& {Roediger}}]{jachym2017}
{J{\'a}chym}, P., {Sun}, M., {Kenney}, J. D.~P., {et~al.} 2017, ApJ, 839, 114

\bibitem[{{Jian} {et~al.}(2018){Jian}, {Lin}, {Oguri}, {Nishizawa}, {Takada},
  {More}, {Koyama}, {Tanaka}, \& {Komiyama}}]{jian2018}
{Jian}, H.-Y., {Lin}, L., {Oguri}, M., {et~al.} 2018, PASJ, 70, S23

\bibitem[{{Jim{\'e}nez-Donaire} {et~al.}(2023){Jim{\'e}nez-Donaire}, {Brown},
  {Wilson}, {Roberts}, {Zabel}, {Ellison}, {Thorp}, {Villanueva}, {Chown},
  {Bisaria}, {Bolatto}, {Boselli}, {Catinella}, {Chung}, {Cortese}, {Davis},
  {Lagos}, {Lee}, {Parker}, {Spekkens}, {Stevens}, \&
  {Sun}}]{jimenez-donaire2022}
{Jim{\'e}nez-Donaire}, M.~J., {Brown}, T., {Wilson}, C.~D., {et~al.} 2023,
  A\&A, 671, A3

\bibitem[{{Kenney} \& {Young}(1986)}]{kenney1986}
{Kenney}, J.~D. \& {Young}, J.~S. 1986, ApJL, 301, L13

\bibitem[{{Kenney} \& {Young}(1988)}]{kenney1988}
{Kenney}, J.~D. \& {Young}, J.~S. 1988, ApJS, 66, 261

\bibitem[{{Kenney} {et~al.}(2015){Kenney}, {Abramson}, \&
  {Bravo-Alfaro}}]{kenney2015}
{Kenney}, J.~D.~P., {Abramson}, A., \& {Bravo-Alfaro}, H. 2015, AJ, 150, 59

\bibitem[{{Kenney} {et~al.}(2014){Kenney}, {Geha}, {J{\'a}chym}, {Crowl},
  {Dague}, {Chung}, {van Gorkom}, \& {Vollmer}}]{kenney2014}
{Kenney}, J. D.~P., {Geha}, M., {J{\'a}chym}, P., {et~al.} 2014, ApJ, 780, 119

\bibitem[{{Kenney} {et~al.}(2004){Kenney}, {van Gorkom}, \&
  {Vollmer}}]{kenney2004}
{Kenney}, J.~D.~P., {van Gorkom}, J.~H., \& {Vollmer}, B. 2004, AJ, 127, 3361

\bibitem[{{Kennicutt}(1998)}]{kennicutt1998}
{Kennicutt}, Robert~C., J. 1998, ApJ, 498, 541

\bibitem[{{Kennicutt} {et~al.}(2003){Kennicutt}, {Armus}, {Bendo}, {Calzetti},
  {Dale}, {Draine}, {Engelbracht}, {Gordon}, {Grauer}, {Helou}, {Hollenbach},
  {Jarrett}, {Kewley}, {Leitherer}, {Li}, {Malhotra}, {Regan}, {Rieke},
  {Rieke}, {Roussel}, {Smith}, {Thornley}, \& {Walter}}]{kennicutt2003}
{Kennicutt}, Robert~C., J., {Armus}, L., {Bendo}, G., {et~al.} 2003, PASP, 115,
  928

\bibitem[{{Kleiner} {et~al.}(2021){Kleiner}, {Serra}, {Maccagni}, {Venhola},
  {Morokuma-Matsui}, {Peletier}, {Iodice}, {Raj}, {de Blok}, {Comrie},
  {J{\'o}zsa}, {Kamphuis}, {Loni}, {Loubser}, {Moln{\'a}r}, {Passmoor},
  {Ramatsoku}, {Sivitilli}, {Smirnov}, {Thorat}, \& {Vitello}}]{kleiner2021}
{Kleiner}, D., {Serra}, P., {Maccagni}, F.~M., {et~al.} 2021, A\&A, 648, A32

\bibitem[{{Knuth}(2006)}]{knuth2006}
{Knuth}, K.~H. 2006, arXiv e-prints, physics/0605197

\bibitem[{{Koopmann} \& {Kenney}(2004)}]{koopmann2004}
{Koopmann}, R.~A. \& {Kenney}, J. D.~P. 2004, ApJ, 613, 866

\bibitem[{{Koribalski} {et~al.}(2018){Koribalski}, {Wang}, {Kamphuis},
  {Westmeier}, {Staveley-Smith}, {Oh}, {L{\'o}pez-S{\'a}nchez}, {Wong}, {Ott},
  {de Blok}, \& {Shao}}]{koribalski2018}
{Koribalski}, B.~S., {Wang}, J., {Kamphuis}, P., {et~al.} 2018, MNRAS, 478,
  1611

\bibitem[{{Lee} \& {Chung}(2018)}]{lee2018}
{Lee}, B. \& {Chung}, A. 2018, ApJL, 866, L10

\bibitem[{{Lee} {et~al.}(2017){Lee}, {Chung}, {Tonnesen}, {Kenney}, {Wong},
  {Vollmer}, {Petitpas}, {Crowl}, \& {van Gorkom}}]{lee2017}
{Lee}, B., {Chung}, A., {Tonnesen}, S., {et~al.} 2017, MNRAS, 466, 1382

\bibitem[{{Lee} {et~al.}(2022{\natexlab{a}}){Lee}, {Wang}, {Chung}, {Ho},
  {Wang}, {Michiyama}, {Molina}, {Kim}, {Shao}, {Kilborn}, {Wang}, {Lin},
  {Kim}, {Catinella}, {Cortese}, {Deg}, {Denes}, {Elagali}, {For}, {Kleiner},
  {Koribalski}, {Lee-Waddell}, {Rhee}, {Spekkens}, {Westmeier}, {Wong},
  {Bigiel}, {Bosma}, {Holwerda}, {van der Hulst}, {Roychowdhury},
  {Verdes-Montenegro}, \& {Zwaan}}]{lee2022}
{Lee}, B., {Wang}, J., {Chung}, A., {et~al.} 2022{\natexlab{a}}, ApJS, 262, 31

\bibitem[{{Lee} {et~al.}(2022{\natexlab{b}}){Lee}, {Lee}, {Mun}, {Cho}, \&
  {Kang}}]{lee2022_enhanced_sfr}
{Lee}, J.~H., {Lee}, M.~G., {Mun}, J.~Y., {Cho}, B.~S., \& {Kang}, J.
  2022{\natexlab{b}}, ApJL, 931, L22

\bibitem[{{Lee} \& {Jang}(2016)}]{lee2016}
{Lee}, M.~G. \& {Jang}, I.~S. 2016, ApJ, 819, 77

\bibitem[{{Lee} {et~al.}(2022{\natexlab{c}}){Lee}, {Sheen}, {Yoon},
  {Jaff{\'e}}, \& {Chung}}]{lee2022_virgo_tails}
{Lee}, S., {Sheen}, Y.-K., {Yoon}, H., {Jaff{\'e}}, Y., \& {Chung}, A.
  2022{\natexlab{c}}, MNRAS, 517, 2912

\bibitem[{{Lelli} {et~al.}(2014){Lelli}, {Verheijen}, \&
  {Fraternali}}]{lelli2014}
{Lelli}, F., {Verheijen}, M., \& {Fraternali}, F. 2014, MNRAS, 445, 1694

\bibitem[{{Leroy} {et~al.}(2021{\natexlab{a}}){Leroy}, {Hughes}, {Liu}, {Pety},
  {Rosolowsky}, {Saito}, {Schinnerer}, {Schruba}, {Usero}, {Faesi}, {Herrera},
  {Chevance}, {Hygate}, {Kepley}, {Koch}, {Querejeta}, {Sliwa}, {Will},
  {Wilson}, {Anand}, {Barnes}, {Belfiore}, {Be{\v{s}}li{\'c}}, {Bigiel},
  {Blanc}, {Bolatto}, {Boquien}, {Cao}, {Chandar}, {Chastenet}, {Chiang},
  {Congiu}, {Dale}, {Deger}, {den Brok}, {Eibensteiner}, {Emsellem},
  {Garc{\'\i}a-Rodr{\'\i}guez}, {Glover}, {Grasha}, {Groves}, {Henshaw},
  {Jim{\'e}nez Donaire}, {Kim}, {Klessen}, {Kreckel}, {Kruijssen}, {Larson},
  {Lee}, {Mayker}, {McElroy}, {Meidt}, {Mok}, {Pan}, {Puschnig}, {Razza},
  {S{\'a}nchez-Bl'azquez}, {Sandstrom}, {Santoro}, {Sardone}, {Scheuermann},
  {Sun}, {Thilker}, {Turner}, {Ubeda}, {Utomo}, {Watkins}, \&
  {Williams}}]{leroy2021_pipeline}
{Leroy}, A.~K., {Hughes}, A., {Liu}, D., {et~al.} 2021{\natexlab{a}}, \apjs,
  255, 19

\bibitem[{{Leroy} {et~al.}(2019){Leroy}, {Sandstrom}, {Lang}, {Lewis}, {Salim},
  {Behrens}, {Chastenet}, {Chiang}, {Gallagher}, {Kessler}, \&
  {Utomo}}]{leroy2019}
{Leroy}, A.~K., {Sandstrom}, K.~M., {Lang}, D., {et~al.} 2019, ApJS, 244, 24

\bibitem[{{Leroy} {et~al.}(2021{\natexlab{b}}){Leroy}, {Schinnerer}, {Hughes},
  {Rosolowsky}, {Pety}, {Schruba}, {Usero}, {Blanc}, {Chevance}, {Emsellem},
  {Faesi}, {Herrera}, {Liu}, {Meidt}, {Querejeta}, {Saito}, {Sandstrom}, {Sun},
  {Williams}, {Anand}, {Barnes}, {Behrens}, {Belfiore}, {Benincasa},
  {Be{\v{s}}li{\'c}}, {Bigiel}, {Bolatto}, {den Brok}, {Cao}, {Chandar},
  {Chastenet}, {Chiang}, {Congiu}, {Dale}, {Deger}, {Eibensteiner}, {Egorov},
  {Garc{\'\i}a-Rodr{\'\i}guez}, {Glover}, {Grasha}, {Henshaw}, {Ho}, {Kepley},
  {Kim}, {Klessen}, {Kreckel}, {Koch}, {Kruijssen}, {Larson}, {Lee}, {Lopez},
  {Machado}, {Mayker}, {McElroy}, {Murphy}, {Ostriker}, {Pan}, {Pessa},
  {Puschnig}, {Razza}, {S{\'a}nchez-Bl{\'a}zquez}, {Santoro}, {Sardone},
  {Scheuermann}, {Sliwa}, {Sormani}, {Stuber}, {Thilker}, {Turner}, {Utomo},
  {Watkins}, \& {Whitmore}}]{leroy2021_phangs}
{Leroy}, A.~K., {Schinnerer}, E., {Hughes}, A., {et~al.} 2021{\natexlab{b}},
  ApJS, 257, 43

\bibitem[{{Leroy} {et~al.}(2009){Leroy}, {Walter}, {Bigiel}, {Usero}, {Weiss},
  {Brinks}, {de Blok}, {Kennicutt}, {Schuster}, {Kramer}, {Wiesemeyer}, \&
  {Roussel}}]{leroy2009}
{Leroy}, A.~K., {Walter}, F., {Bigiel}, F., {et~al.} 2009, AJ, 137, 4670

\bibitem[{{Lin} {et~al.}(2014){Lin}, {Jian}, {Foucaud}, {Norberg}, {Bower},
  {Cole}, {Arnalte-Mur}, {Chen}, {Coupon}, {Hsieh}, {Heinis}, {Phleps}, {Chen},
  {Lee}, {Burgett}, {Chambers}, {Denneau}, {Draper}, {Flewelling}, {Hodapp},
  {Huber}, {Kaiser}, {Kudritzki}, {Magnier}, {Metcalfe}, {Price}, {Tonry},
  {Wainscoat}, \& {Waters}}]{lin2014}
{Lin}, L., {Jian}, H.-Y., {Foucaud}, S., {et~al.} 2014, ApJ, 782, 33

\bibitem[{{Lisenfeld} {et~al.}(2016){Lisenfeld}, {Braine}, {Duc}, {Boquien},
  {Brinks}, {Bournaud}, {Lelli}, \& {Charmandaris}}]{lisenfeld2016}
{Lisenfeld}, U., {Braine}, J., {Duc}, P.~A., {et~al.} 2016, A\&A, 590, A92

\bibitem[{{Liz{\'e}e} {et~al.}(2021){Liz{\'e}e}, {Vollmer}, {Braine}, \&
  {Nehlig}}]{lizee2021}
{Liz{\'e}e}, T., {Vollmer}, B., {Braine}, J., \& {Nehlig}, F. 2021, A\&A, 645,
  A111

\bibitem[{{Loni} {et~al.}(2021){Loni}, {Serra}, {Kleiner}, {Cortese},
  {Catinella}, {Koribalski}, {Jarrett}, {Molnar}, {Davis}, {Iodice},
  {Lee-Waddell}, {Loi}, {Maccagni}, {Peletier}, {Popping}, {Ramatsoku},
  {Smith}, \& {Zabel}}]{loni2021}
{Loni}, A., {Serra}, P., {Kleiner}, D., {et~al.} 2021, A\&A, 648, A31

\bibitem[{{Makarov} {et~al.}(2014){Makarov}, {Prugniel}, {Terekhova},
  {Courtois}, \& {Vauglin}}]{makarov2014}
{Makarov}, D., {Prugniel}, P., {Terekhova}, N., {Courtois}, H., \& {Vauglin},
  I. 2014, A\&A, 570, A13

\bibitem[{{Matthews} {et~al.}(1998){Matthews}, {van Driel}, \&
  {Gallagher}}]{matthews1998}
{Matthews}, L.~D., {van Driel}, W., \& {Gallagher}, J.~S., I. 1998, AJ, 116,
  1169

\bibitem[{{Mayer} {et~al.}(2006){Mayer}, {Mastropietro}, {Wadsley}, {Stadel},
  \& {Moore}}]{mayer2006}
{Mayer}, L., {Mastropietro}, C., {Wadsley}, J., {Stadel}, J., \& {Moore}, B.
  2006, MNRAS, 369, 1021

\bibitem[{{Mei} {et~al.}(2007){Mei}, {Blakeslee}, {C{\^o}t{\'e}}, {Tonry},
  {West}, {Ferrarese}, {Jord{\'a}n}, {Peng}, {Anthony}, \& {Merritt}}]{mei2007}
{Mei}, S., {Blakeslee}, J.~P., {C{\^o}t{\'e}}, P., {et~al.} 2007, ApJ, 655, 144

\bibitem[{{Meyer} {et~al.}(2017){Meyer}, {Robotham}, {Obreschkow}, {Westmeier},
  {Duffy}, \& {Staveley-Smith}}]{meyer2017}
{Meyer}, M., {Robotham}, A., {Obreschkow}, D., {et~al.} 2017, PASA, 34, 52

\bibitem[{{Mok} {et~al.}(2016){Mok}, {Wilson}, {Golding}, {Warren}, {Israel},
  {Serjeant}, {Knapen}, {S{\'a}nchez-Gallego}, {Barmby}, {Bendo}, {Rosolowsky},
  \& {van der Werf}}]{mok2016}
{Mok}, A., {Wilson}, C.~D., {Golding}, J., {et~al.} 2016, MNRAS, 456, 4384

\bibitem[{{Mok} {et~al.}(2017){Mok}, {Wilson}, {Knapen}, {S{\'a}nchez-Gallego},
  {Brinks}, \& {Rosolowsky}}]{mok2017}
{Mok}, A., {Wilson}, C.~D., {Knapen}, J.~H., {et~al.} 2017, MNRAS, 467, 4282

\bibitem[{{Moln{\'a}r} {et~al.}(2022){Moln{\'a}r}, {Serra}, {van der Hulst},
  {Jarrett}, {Boselli}, {Cortese}, {Healy}, {de Blok}, {Cappellari}, {Hess},
  {J{\'o}zsa}, {McDermid}, {Oosterloo}, \& {Verheijen}}]{molnar2022}
{Moln{\'a}r}, D.~C., {Serra}, P., {van der Hulst}, T., {et~al.} 2022, A\&A,
  659, A94

\bibitem[{{Moore} {et~al.}(1996){Moore}, {Katz}, {Lake}, {Dressler}, \&
  {Oemler}}]{moore1996}
{Moore}, B., {Katz}, N., {Lake}, G., {Dressler}, A., \& {Oemler}, A. 1996,
  Nature, 379, 613

\bibitem[{{Moretti} {et~al.}(2018){Moretti}, {Paladino}, {Poggianti},
  {D'Onofrio}, {Bettoni}, {Gullieuszik}, {Jaff{\'e}}, {Vulcani}, {Fasano},
  {Fritz}, \& {Torstensson}}]{moretti2018}
{Moretti}, A., {Paladino}, R., {Poggianti}, B.~M., {et~al.} 2018, MNRAS, 480,
  2508

\bibitem[{{Moretti} {et~al.}(2020{\natexlab{a}}){Moretti}, {Paladino},
  {Poggianti}, {Serra}, {Ramatsoku}, {Franchetto}, {Deb}, {Gullieuszik},
  {Tomi{\v{c}}i{\'c}}, {Mingozzi}, {Vulcani}, {Radovich}, {Bettoni}, \&
  {Fritz}}]{moretti2020b}
{Moretti}, A., {Paladino}, R., {Poggianti}, B.~M., {et~al.} 2020{\natexlab{a}},
  ApJL, 897, L30

\bibitem[{{Moretti} {et~al.}(2020{\natexlab{b}}){Moretti}, {Paladino},
  {Poggianti}, {Serra}, {Roediger}, {Gullieuszik}, {Tomi{\v{c}}i{\'c}},
  {Radovich}, {Vulcani}, {Jaff{\'e}}, {Fritz}, {Bettoni}, {Ramatsoku}, \&
  {Wolter}}]{moretti2020}
{Moretti}, A., {Paladino}, R., {Poggianti}, B.~M., {et~al.} 2020{\natexlab{b}},
  ApJ, 889, 9

\bibitem[{{Morokuma-Matsui} {et~al.}(2022){Morokuma-Matsui}, {Bekki}, {Wang},
  {Serra}, {Koyama}, {Morokuma}, {Egusa}, {For}, {Nakanishi}, {Koribalski},
  {Okamoto}, {Kodama}, {Lee}, {Maccagni}, {Miura}, {Espada}, {Takeuchi},
  {Yang}, {Lee}, {Ueda}, \& {Matsushita}}]{morokuma-matsui2022}
{Morokuma-Matsui}, K., {Bekki}, K., {Wang}, J., {et~al.} 2022, ApJS, 263, 40

\bibitem[{{Nehlig} {et~al.}(2016){Nehlig}, {Vollmer}, \& {Braine}}]{nehlig2016}
{Nehlig}, F., {Vollmer}, B., \& {Braine}, J. 2016, A\&A, 587, A108

\bibitem[{{Nelson} {et~al.}(2019){Nelson}, {Springel}, {Pillepich},
  {Rodriguez-Gomez}, {Torrey}, {Genel}, {Vogelsberger}, {Pakmor}, {Marinacci},
  {Weinberger}, {Kelley}, {Lovell}, {Diemer}, \& {Hernquist}}]{nelson2019}
{Nelson}, D., {Springel}, V., {Pillepich}, A., {et~al.} 2019, Computational
  Astrophysics and Cosmology, 6, 2

\bibitem[{{Oosterloo} \& {van Gorkom}(2005)}]{oosterloo2005}
{Oosterloo}, T. \& {van Gorkom}, J. 2005, A\&A, 437, L19

\bibitem[{{Pappalardo} {et~al.}(2010){Pappalardo}, {Lan{\c{c}}on}, {Vollmer},
  {Ocvirk}, {Boissier}, \& {Boselli}}]{pappalardo2010}
{Pappalardo}, C., {Lan{\c{c}}on}, A., {Vollmer}, B., {et~al.} 2010, A\&A, 514,
  A33

\bibitem[{{Peng} {et~al.}(2010){Peng}, {Lilly}, {Kova{\v c}}, {Bolzonella},
  {Pozzetti}, {Renzini}, {Zamorani}, {Ilbert}, {Knobel}, {Iovino}, {Maier},
  {Cucciati}, {Tasca}, {Carollo}, {Silverman}, {Kampczyk}, {de Ravel},
  {Sanders}, {Scoville}, {Contini}, {Mainieri}, {Scodeggio}, {Kneib}, {Le
  F{\`e}vre}, {Bardelli}, {Bongiorno}, {Caputi}, {Coppa}, {de la Torre},
  {Franzetti}, {Garilli}, {Lamareille}, {Le Borgne}, {Le Brun}, {Mignoli},
  {Perez Montero}, {Pello}, {Ricciardelli}, {Tanaka}, {Tresse}, {Vergani},
  {Welikala}, {Zucca}, {Oesch}, {Abbas}, {Barnes}, {Bordoloi}, {Bottini},
  {Cappi}, {Cassata}, {Cimatti}, {Fumana}, {Hasinger}, {Koekemoer},
  {Leauthaud}, {Maccagni}, {Marinoni}, {McCracken}, {Memeo}, {Meneux}, {Nair},
  {Porciani}, {Presotto}, \& {Scaramella}}]{peng2010}
{Peng}, Y.-j., {Lilly}, S.~J., {Kova{\v c}}, K., {et~al.} 2010, ApJ, 721, 193

\bibitem[{{Pillepich} {et~al.}(2018){Pillepich}, {Springel}, {Nelson}, {Genel},
  {Naiman}, {Pakmor}, {Hernquist}, {Torrey}, {Vogelsberger}, {Weinberger}, \&
  {Marinacci}}]{pillepich2018}
{Pillepich}, A., {Springel}, V., {Nelson}, D., {et~al.} 2018, MNRAS, 473, 4077

\bibitem[{{Poggianti} {et~al.}(2017){Poggianti}, {Moretti}, {Gullieuszik},
  {Fritz}, {Jaff{\'e}}, {Bettoni}, {Fasano}, {Bellhouse}, {Hau}, {Vulcani},
  {Biviano}, {Omizzolo}, {Paccagnella}, {D'Onofrio}, {Cava}, {Sheen}, {Couch},
  \& {Owers}}]{poggianti2017}
{Poggianti}, B.~M., {Moretti}, A., {Gullieuszik}, M., {et~al.} 2017, ApJ, 844,
  48

\bibitem[{{Postman} \& {Geller}(1984)}]{postman1984}
{Postman}, M. \& {Geller}, M.~J. 1984, ApJ, 281, 95

\bibitem[{{Quilis} {et~al.}(2000){Quilis}, {Moore}, \& {Bower}}]{quilis2000}
{Quilis}, V., {Moore}, B., \& {Bower}, R. 2000, Science, 288, 1617

\bibitem[{{Rasmussen} {et~al.}(2006){Rasmussen}, {Ponman}, \&
  {Mulchaey}}]{rasmussen2006}
{Rasmussen}, J., {Ponman}, T.~J., \& {Mulchaey}, J.~S. 2006, MNRAS, 370, 453

\bibitem[{{Reynolds} {et~al.}(2020){Reynolds}, {Westmeier}, {Staveley-Smith},
  {Chauhan}, \& {Lagos}}]{reynolds2020}
{Reynolds}, T.~N., {Westmeier}, T., {Staveley-Smith}, L., {Chauhan}, G., \&
  {Lagos}, C.~D.~P. 2020, MNRAS, 493, 5089

\bibitem[{{Richter} \& {Sancisi}(1994)}]{richter1994}
{Richter}, O.~G. \& {Sancisi}, R. 1994, A\&A, 290, L9

\bibitem[{{Roberts} {et~al.}(2022{\natexlab{a}}){Roberts}, {Lang}, {Trotsenko},
  {Bemis}, {Ellison}, {Lin}, {Pan}, {Ignesti}, {Leslie}, \& {van
  Weeren}}]{roberts2022_lofar_manga}
{Roberts}, I.~D., {Lang}, M., {Trotsenko}, D., {et~al.} 2022{\natexlab{a}},
  ApJ, 941, 77

\bibitem[{{Roberts} \& {Parker}(2020)}]{roberts2020}
{Roberts}, I.~D. \& {Parker}, L.~C. 2020, MNRAS, 495, 554

\bibitem[{{Roberts} {et~al.}(2019){Roberts}, {Parker}, {Brown}, {Joshi},
  {Hlavacek-Larrondo}, \& {Wadsley}}]{roberts2019}
{Roberts}, I.~D., {Parker}, L.~C., {Brown}, T., {et~al.} 2019, ApJ, 873, 42

\bibitem[{{Roberts} {et~al.}(2022{\natexlab{b}}){Roberts}, {Parker}, {Gwyn},
  {Hudson}, {Carlberg}, {McConnachie}, {Cuillandre}, {Chambers}, {Duc},
  {Furusawa}, {Gavazzi}, {Hill}, {Huber}, {Ibata}, {Kilbinger}, {Mei},
  {Mellier}, {Miyazaki}, {Oguri}, \& {Wainscoat}}]{roberts2022_UNIONS}
{Roberts}, I.~D., {Parker}, L.~C., {Gwyn}, S., {et~al.} 2022{\natexlab{b}},
  MNRAS, 509, 1342

\bibitem[{{Roberts} {et~al.}(2021{\natexlab{a}}){Roberts}, {van Weeren},
  {McGee}, {Botteon}, {Drabent}, {Ignesti}, {Rottgering}, {Shimwell}, \&
  {Tasse}}]{roberts2021_LOFARclust}
{Roberts}, I.~D., {van Weeren}, R.~J., {McGee}, S.~L., {et~al.}
  2021{\natexlab{a}}, \aap, 650, A111

\bibitem[{{Roberts} {et~al.}(2021{\natexlab{b}}){Roberts}, {van Weeren},
  {McGee}, {Botteon}, {Ignesti}, \& {Rottgering}}]{roberts2021_LOFARgrp}
{Roberts}, I.~D., {van Weeren}, R.~J., {McGee}, S.~L., {et~al.}
  2021{\natexlab{b}}, A\&A, 652, A153

\bibitem[{{Roberts} {et~al.}(2022{\natexlab{c}}){Roberts}, {van Weeren},
  {Timmerman}, {Botteon}, {Gendron-Marsolais}, {Ignesti}, \&
  {Rottgering}}]{roberts2022_perseus}
{Roberts}, I.~D., {van Weeren}, R.~J., {Timmerman}, R., {et~al.}
  2022{\natexlab{c}}, A\&A, 658, A44

\bibitem[{{Saintonge} {et~al.}(2017){Saintonge}, {Catinella}, {Tacconi},
  {Kauffmann}, {Genzel}, {Cortese}, {Dav{\'e}}, {Fletcher},
  {Graci{\'a}-Carpio}, {Kramer}, {Heckman}, {Janowiecki}, {Lutz}, {Rosario},
  {Schiminovich}, {Schuster}, {Wang}, {Wuyts}, {Borthakur}, {Lamperti}, \&
  {Roberts-Borsani}}]{saintonge2017}
{Saintonge}, A., {Catinella}, B., {Tacconi}, L.~J., {et~al.} 2017, ApJS, 233,
  22

\bibitem[{{Schade} {et~al.}(1995){Schade}, {Lilly}, {Crampton}, {Hammer}, {Le
  Fevre}, \& {Tresse}}]{schade1995}
{Schade}, D., {Lilly}, S.~J., {Crampton}, D., {et~al.} 1995, ApJL, 451, L1

\bibitem[{{Schaefer} {et~al.}(2017){Schaefer}, {Croom}, {Allen}, {Brough},
  {Medling}, {Ho}, {Scott}, {Richards}, {Pracy}, {Gunawardhana}, {Norberg},
  {Alpaslan}, {Bauer}, {Bekki}, {Bland-Hawthorn}, {Bloom}, {Bryant}, {Couch},
  {Driver}, {Fogarty}, {Foster}, {Goldstein}, {Green}, {Hopkins},
  {Konstantopoulos}, {Lawrence}, {L{\'o}pez-S{\'a}nchez}, {Lorente}, {Owers},
  {Sharp}, {Sweet}, {Taylor}, {van de Sande}, {Walcher}, \&
  {Wong}}]{schaefer2017}
{Schaefer}, A.~L., {Croom}, S.~M., {Allen}, J.~T., {et~al.} 2017, MNRAS, 464,
  121

\bibitem[{{Schaefer} {et~al.}(2019){Schaefer}, {Croom}, {Scott}, {Brough},
  {Allen}, {Bekki}, {Bland-Hawthorn}, {Bloom}, {Bryant}, {Cortese}, {Davies},
  {Federrath}, {Fogarty}, {Green}, {Groves}, {Hopkins}, {Konstantopoulos},
  {L{\'o}pez-S{\'a}nchez}, {Lawrence}, {McElroy}, {Medling}, {Owers}, {Pracy},
  {Richards}, {Robotham}, {van de Sande}, {Tonini}, \& {Yi}}]{schaefer2019}
{Schaefer}, A.~L., {Croom}, S.~M., {Scott}, N., {et~al.} 2019, MNRAS, 483, 2851

\bibitem[{{Schmidt}(1959)}]{schmidt1959}
{Schmidt}, M. 1959, ApJ, 129, 243

\bibitem[{Scholz \& Stephens(1987)}]{scholz1987}
Scholz, F.~W. \& Stephens, M.~A. 1987, Journal of the American Statistical
  Association, 82, 918

\bibitem[{{Solomon} {et~al.}(1987){Solomon}, {Rivolo}, {Barrett}, \&
  {Yahil}}]{solomon1987}
{Solomon}, P.~M., {Rivolo}, A.~R., {Barrett}, J., \& {Yahil}, A. 1987, ApJ,
  319, 730

\bibitem[{{Steinhauser} {et~al.}(2012){Steinhauser}, {Haider}, {Kapferer}, \&
  {Schindler}}]{steinhauser2012}
{Steinhauser}, D., {Haider}, M., {Kapferer}, W., \& {Schindler}, S. 2012, A\&A,
  544, A54

\bibitem[{{Stevens} \& {Brown}(2017)}]{stevens2017}
{Stevens}, A. R.~H. \& {Brown}, T. 2017, MNRAS, 471, 447

\bibitem[{{Stevens} {et~al.}(2021){Stevens}, {Lagos}, {Cortese}, {Catinella},
  {Diemer}, {Nelson}, {Pillepich}, {Hernquist}, {Marinacci}, \&
  {Vogelsberger}}]{stevens2021}
{Stevens}, A. R.~H., {Lagos}, C. d.~P., {Cortese}, L., {et~al.} 2021, MNRAS,
  502, 3158

\bibitem[{{Thorp} {et~al.}(2021){Thorp}, {Bluck}, {Ellison}, {Maiolino},
  {Conselice}, {Hani}, \& {Bottrell}}]{thorp2021}
{Thorp}, M.~D., {Bluck}, A. F.~L., {Ellison}, S.~L., {et~al.} 2021, MNRAS, 507,
  886

\bibitem[{{Tonnesen} \& {Bryan}(2009)}]{tonnesen2009}
{Tonnesen}, S. \& {Bryan}, G.~L. 2009, ApJ, 694, 789

\bibitem[{{Troncoso-Iribarren} {et~al.}(2020){Troncoso-Iribarren}, {Padilla},
  {Santander}, {Lagos}, {Garc{\'\i}a-Lambas}, {Rodr{\'\i}guez}, \&
  {Contreras}}]{troncoso-iribarren2020}
{Troncoso-Iribarren}, P., {Padilla}, N., {Santander}, C., {et~al.} 2020, MNRAS,
  497, 4145

\bibitem[{{Verdugo} {et~al.}(2015){Verdugo}, {Combes}, {Dasyra}, {Salom{\'e}},
  \& {Braine}}]{verdugo2015}
{Verdugo}, C., {Combes}, F., {Dasyra}, K., {Salom{\'e}}, P., \& {Braine}, J.
  2015, A\&A, 582, A6

\bibitem[{{Villanueva} {et~al.}(2022){Villanueva}, {Bolatto}, {Vogel}, {Brown},
  {Wilson}, {Zabel}, {Ellison}, {Stevens}, {Jim{\'e}nez Donaire}, {Spekkens},
  {Tharp}, {Davis}, {Parker}, {Roberts}, {Basra}, {Boselli}, {Catinella},
  {Chung}, {Cortese}, {Lee}, \& {Watts}}]{villanueva2022}
{Villanueva}, V., {Bolatto}, A.~D., {Vogel}, S., {et~al.} 2022, ApJ, 940, 176

\bibitem[{{Vollmer} {et~al.}(2008){Vollmer}, {Braine}, {Pappalardo}, \&
  {Hily-Blant}}]{vollmer2008}
{Vollmer}, B., {Braine}, J., {Pappalardo}, C., \& {Hily-Blant}, P. 2008, A\&A,
  491, 455

\bibitem[{{Vollmer} {et~al.}(2012){Vollmer}, {Soida}, {Braine}, {Abramson},
  {Beck}, {Chung}, {Crowl}, {Kenney}, \& {van Gorkom}}]{vollmer2012}
{Vollmer}, B., {Soida}, M., {Braine}, J., {et~al.} 2012, A\&A, 537, A143

\bibitem[{{Vollmer} {et~al.}(2009){Vollmer}, {Soida}, {Chung}, {Chemin},
  {Braine}, {Boselli}, \& {Beck}}]{vollmer2009}
{Vollmer}, B., {Soida}, M., {Chung}, A., {et~al.} 2009, A\&A, 496, 669

\bibitem[{{Vulcani} {et~al.}(2018{\natexlab{a}}){Vulcani}, {Poggianti},
  {Gullieuszik}, {Moretti}, {Tonnesen}, {Jaff{\'e}}, {Fritz}, {Fasano}, \&
  {Bettoni}}]{vulcani2018_sf}
{Vulcani}, B., {Poggianti}, B.~M., {Gullieuszik}, M., {et~al.}
  2018{\natexlab{a}}, ApJL, 866, L25

\bibitem[{{Vulcani} {et~al.}(2018{\natexlab{b}}){Vulcani}, {Poggianti},
  {Jaff{\'e}}, {Moretti}, {Fritz}, {Gullieuszik}, {Bettoni}, {Fasano},
  {Tonnesen}, \& {McGee}}]{vulcani2018_grp}
{Vulcani}, B., {Poggianti}, B.~M., {Jaff{\'e}}, Y.~L., {et~al.}
  2018{\natexlab{b}}, MNRAS, 480, 3152

\bibitem[{{Vulcani} {et~al.}(2020){Vulcani}, {Poggianti}, {Tonnesen}, {McGee},
  {Moretti}, {Fritz}, {Gullieuszik}, {Jaff{\'e}}, {Franchetto},
  {Tomi{\v{c}}i{\'c}}, {Mingozzi}, {Bettoni}, \& {Wolter}}]{vulcani2020}
{Vulcani}, B., {Poggianti}, B.~M., {Tonnesen}, S., {et~al.} 2020, ApJ, 899, 98

\bibitem[{{Walter} {et~al.}(2008){Walter}, {Brinks}, {de Blok}, {Bigiel},
  {Kennicutt}, {Thornley}, \& {Leroy}}]{walter2008}
{Walter}, F., {Brinks}, E., {de Blok}, W.~J.~G., {et~al.} 2008, AJ, 136, 2563

\bibitem[{{Watts} {et~al.}(2020{\natexlab{a}}){Watts}, {Catinella}, {Cortese},
  \& {Power}}]{watts2020a}
{Watts}, A.~B., {Catinella}, B., {Cortese}, L., \& {Power}, C.
  2020{\natexlab{a}}, MNRAS, 492, 3672

\bibitem[{{Watts} {et~al.}(2023){Watts}, {Cortese}, {Catinella}, {Brown},
  {Wilson}, {Zabel}, {Roberts}, {Davis}, {Thorp}, {Chung}, {Stevens},
  {Ellison}, {Spekkens}, {Parker}, {Bah{\'e}}, {Villanueva},
  {Jim{\'e}nez-Donaire}, {Bisaria}, {Boselli}, {Bolatto}, \& {Lee}}]{watts2023}
{Watts}, A.~B., {Cortese}, L., {Catinella}, B., {et~al.} 2023, arXiv e-prints,
  arXiv:2303.07549

\bibitem[{{Watts} {et~al.}(2020{\natexlab{b}}){Watts}, {Power}, {Catinella},
  {Cortese}, \& {Stevens}}]{watts2020b}
{Watts}, A.~B., {Power}, C., {Catinella}, B., {Cortese}, L., \& {Stevens}, A.
  R.~H. 2020{\natexlab{b}}, MNRAS, 499, 5205

\bibitem[{{Wetzel} {et~al.}(2012){Wetzel}, {Tinker}, \& {Conroy}}]{wetzel2012}
{Wetzel}, A.~R., {Tinker}, J.~L., \& {Conroy}, C. 2012, MNRAS, 424, 232

\bibitem[{{Yagi} {et~al.}(2010){Yagi}, {Yoshida}, {Komiyama}, {Kashikawa},
  {Furusawa}, {Okamura}, {Graham}, {Miller}, {Carter}, {Mobasher}, \&
  {Jogee}}]{yagi2010}
{Yagi}, M., {Yoshida}, M., {Komiyama}, Y., {et~al.} 2010, AJ, 140, 1814

\bibitem[{Yamagami \& Fujita(2011)}]{yamagami2011}
Yamagami, T. \& Fujita, Y. 2011, PASJ, 63, 1165

\bibitem[{{Yoon} {et~al.}(2017){Yoon}, {Chung}, {Smith}, \&
  {Jaff{\'e}}}]{yoon2017}
{Yoon}, H., {Chung}, A., {Smith}, R., \& {Jaff{\'e}}, Y.~L. 2017, ApJ, 838, 81

\bibitem[{{York} {et~al.}(2000){York}, {Adelman}, {Anderson}, {Anderson},
  {Annis}, {Bahcall}, {Bakken}, {Barkhouser}, {Bastian}, {Berman}, {Boroski},
  {Bracker}, {Briegel}, {Briggs}, {Brinkmann}, {Brunner}, {Burles}, {Carey},
  {Carr}, {Castander}, {Chen}, {Colestock}, {Connolly}, {Crocker}, {Csabai},
  {Czarapata}, {Davis}, {Doi}, {Dombeck}, {Eisenstein}, {Ellman}, {Elms},
  {Evans}, {Fan}, {Federwitz}, {Fiscelli}, {Friedman}, {Frieman}, {Fukugita},
  {Gillespie}, {Gunn}, {Gurbani}, {de Haas}, {Haldeman}, {Harris}, {Hayes},
  {Heckman}, {Hennessy}, {Hindsley}, {Holm}, {Holmgren}, {Huang}, {Hull},
  {Husby}, {Ichikawa}, {Ichikawa}, {Ivezi{\'c}}, {Kent}, {Kim}, {Kinney},
  {Klaene}, {Kleinman}, {Kleinman}, {Knapp}, {Korienek}, {Kron}, {Kunszt},
  {Lamb}, {Lee}, {Leger}, {Limmongkol}, {Lindenmeyer}, {Long}, {Loomis},
  {Loveday}, {Lucinio}, {Lupton}, {MacKinnon}, {Mannery}, {Mantsch}, {Margon},
  {McGehee}, {McKay}, {Meiksin}, {Merelli}, {Monet}, {Munn}, {Narayanan},
  {Nash}, {Neilsen}, {Neswold}, {Newberg}, {Nichol}, {Nicinski}, {Nonino},
  {Okada}, {Okamura}, {Ostriker}, {Owen}, {Pauls}, {Peoples}, {Peterson},
  {Petravick}, {Pier}, {Pope}, {Pordes}, {Prosapio}, {Rechenmacher}, {Quinn},
  {Richards}, {Richmond}, {Rivetta}, {Rockosi}, {Ruthmansdorfer}, {Sandford},
  {Schlegel}, {Schneider}, {Sekiguchi}, {Sergey}, {Shimasaku}, {Siegmund},
  {Smee}, {Smith}, {Snedden}, {Stone}, {Stoughton}, {Strauss}, {Stubbs},
  {SubbaRao}, {Szalay}, {Szapudi}, {Szokoly}, {Thakar}, {Tremonti}, {Tucker},
  {Uomoto}, {Vanden Berk}, {Vogeley}, {Waddell}, {Wang}, {Watanabe},
  {Weinberg}, {Yanny}, {Yasuda}, \& {SDSS Collaboration}}]{york2000}
{York}, D.~G., {Adelman}, J., {Anderson}, Jr., J.~E., {et~al.} 2000, AJ, 120,
  1579

\bibitem[{{Yoshida} {et~al.}(2002){Yoshida}, {Yagi}, {Okamura}, {Aoki},
  {Ohyama}, {Komiyama}, {Yasuda}, {Iye}, {Kashikawa}, {Doi}, {Furusawa},
  {Hamabe}, {Kimura}, {Miyazaki}, {Miyazaki}, {Nakata}, {Ouchi}, {Sekiguchi},
  {Shimasaku}, \& {Ohtani}}]{yoshida2002}
{Yoshida}, M., {Yagi}, M., {Okamura}, S., {et~al.} 2002, ApJ, 567, 118

\bibitem[{{Zabel} {et~al.}(2022){Zabel}, {Brown}, {Wilson}, {Davis}, {Cortese},
  {Parker}, {Boselli}, {Catinella}, {Chown}, {Chung}, {Deb}, {Ellison},
  {Jim{\'e}nez-Donaire}, {Lee}, {Roberts}, {Spekkens}, {Stevens}, {Thorp},
  {Tonnesen}, \& {Villanueva}}]{zabel2022}
{Zabel}, N., {Brown}, T., {Wilson}, C.~D., {et~al.} 2022, ApJ, 933, 10

\bibitem[{{Zabel} {et~al.}(2020){Zabel}, {Davis}, {Sarzi}, {Nedelchev},
  {Chevance}, {Kruijssen}, {Iodice}, {Baes}, {Bendo}, {Corsini}, {De Looze},
  {de Zeeuw}, {Gadotti}, {Grossi}, {Peletier}, {Pinna}, {Serra}, {van de
  Voort}, {Venhola}, {Viaene}, \& {Vlahakis}}]{zabel2020}
{Zabel}, N., {Davis}, T.~A., {Sarzi}, M., {et~al.} 2020, MNRAS, 496, 2155

\bibitem[{{Zabel} {et~al.}(2019){Zabel}, {Davis}, {Smith}, {Maddox}, {Bendo},
  {Peletier}, {Iodice}, {Venhola}, {Baes}, {Davies}, {de Looze}, {Gomez},
  {Grossi}, {Kenney}, {Serra}, {van de Voort}, {Vlahakis}, \&
  {Young}}]{zabel2019}
{Zabel}, N., {Davis}, T.~A., {Smith}, M. W.~L., {et~al.} 2019, MNRAS, 483, 2251

\end{thebibliography}

\end{document}